\documentclass[11pt,a4paper]{article}
\pdfoutput=1

\usepackage[includeheadfoot,
            marginratio={1:1,2:3}, 
            width=412pt, 
            height=688pt,]{geometry}

\usepackage[T1]{fontenc}
\usepackage[utf8]{inputenc}
\usepackage[english]{babel}
\usepackage{csquotes}
\usepackage{pifont}

\usepackage{amsmath, amssymb, amsbsy, mathtools}
\usepackage{physics}
\usepackage{booktabs}
\usepackage{multirow}
\usepackage{multicol}
\usepackage{blkarray}

\usepackage[table,dvipsnames]{xcolor}

\usepackage{graphicx}

\usepackage[font=footnotesize, labelfont=bf]{caption}
\usepackage[font=footnotesize, labelfont=bf]{subcaption}
\usepackage[sort&compress, numbers]{natbib}
\usepackage[shortlabels]{enumitem}

\usepackage{lscape}

\usepackage[framemethod=TikZ]{mdframed}

\usepackage{hyperref}

\newcommand{\nc}{\newcommand}
\nc{\lb}{\llbracket}
\nc{\rb}{\rrbracket}
\nc{\gl}{\llbracket}
\nc{\gr}{\rrbracket}
\nc{\del}{\partial}

\allowdisplaybreaks[2]
\numberwithin{equation}{section}

\usepackage{cancel}

\numberwithin{equation}{section}

\def\be{\begin{equation}}
\def\ee{\end{equation}}
\def\bea{\begin{eqnarray}}
\def\eea{\end{eqnarray}}

\mdfdefinestyle{Example}{
    linecolor=black!80!white,
    outerlinewidth=0.5pt,
    roundcorner=0pt,
    innertopmargin=15pt,
    innerbottommargin=15pt,
    innerrightmargin=20pt,
    innerleftmargin=20pt,
    backgroundcolor=yellow!10!white
}
\def\example{\begin{mdframed}[style=Example]}

\mdfdefinestyle{Result}{
    linecolor=black!80!white,
    outerlinewidth=0.5pt,
    roundcorner=2pt,
    innertopmargin=15pt,
    innerbottommargin=15pt,
    innerrightmargin=20pt,
    innerleftmargin=20pt,
    backgroundcolor=blue!5!white
}

\mdfdefinestyle{Question}{
    linecolor=black!80!white,
    outerlinewidth=0.5pt,
    roundcorner=0pt,
    innertopmargin=15pt,
    innerbottommargin=15pt,
    innerrightmargin=20pt,
    innerleftmargin=20pt,
    backgroundcolor=orange!15!white
}
\def\question{\begin{mdframed}[style=Question]}

\mdfdefinestyle{theorem}{
    linecolor=black!80!white,
    outerlinewidth=0.5pt,
    roundcorner=2pt,
    innertopmargin=15pt,
    innerbottommargin=15pt,
    innerrightmargin=20pt,
    innerleftmargin=20pt,
    backgroundcolor=black!10!white
}

\begin{document}

\vspace*{-1.5cm}
\begin{flushright}
  {\small
  IFT-UAM/CSIC-26-45\\
  }
\end{flushright}

\vspace{1.5cm}
\begin{center}
{\LARGE
Sharpened Dynamical Cobordism \\[0.2cm]
} 
\vspace{0.4cm}

\end{center}

\vspace{0.35cm}
\begin{center}
 Andriana Makridou${}^1$, Alejandro Javier Puga Gómez${}^{1,2}$
\end{center}

\vspace{0.1cm}
\begin{center} 
${}^1$\emph{Instituto de Física Teórica, \\ 
Universidad Autónoma de Madrid \& CSIC,\\
C. Nicol{\'a}s Cabrera 13-15, 28049 Cantoblanco, Madrid, Spain} \\[0.3cm]
${}^2$\emph{Departamento de Física Teórica,
Universidad Autónoma de Madrid,\\
28049 Cantoblanco, Madrid, Spain} \\[0.1cm] 
 
 \vspace{0.3cm} 
\end{center} 

\vspace{0.5cm}


\begin{abstract}

We propose a sharpened version of Dynamical Cobordism, where the physical structure $\xi$ of the theory in question determines an allowed range $\mathcal{R}^\xi$ for the critical exponent $\delta$. We interpret a singularity with $\delta \in \mathcal{R}^\xi$ as a true transition-to-nothing, i.e., a configuration ending spacetime, while a singularity with $\delta \notin \mathcal{R}^\xi$ indicates some obstruction to such a transition, i.e., the presence of a non-trivial cobordism global charge, which is incompatible with a theory of quantum gravity. In the spirit of the original Cobordism Conjecture, this apparent inconsistency of the theory can be alleviated via the modification of the structure, for instance by introducing new degrees of freedom and associated defects. Inspired by the Gubser criterion for good singularities, we propose a way to determine $\mathcal{R}^\xi$. As a proof-of-concept we show explicitly how the introduction of a higher-form gauge field changes the allowed range of $\delta$ compared to an EFT with only scalars. We test this sharpened version of Dynamical Cobordism against several examples, such as massive IIA string theory, where it is notably compatible with the presence of O8-planes; the Janis-Newman-Winicour and Garfinkle-Horowitz-Strominger black hole solutions; and certain singular distributions of D-branes. In all these cases, the Sharpened Dynamical Cobordism Conjecture leads to results consistent with our expectations.
\end{abstract}

\clearpage

\tableofcontents



\section{Introduction}

\label{sec:intro}

Within gravitational theories, solutions with spacetime singularities, where some curvature invariant becomes infinite, can arise frequently, especially in setups with a limited amount of supersymmetry, see e.g., \cite{Leone:2025mwo, Dudas:2025ubq}. A lot of effort has been put towards determining whether a spacetime singularity can be viewed as acceptable, possibly signaling some limitation of the effective theory, or whether it signifies some fundamental inconsistency of the solution. For instance, the Weak Cosmic Censorship Conjecture \cite{Penrose:1969} postulates that, in general, the singularities should be hidden behind a black hole horizon, hence any naked singularities should be deemed unphysical.

In a holographic context, in 2000, Gubser classified singularities as \emph{good} or \emph{bad}, depending on whether a horizon can form around them \cite{Gubser:2000nd}. This led to the famous Gubser criterion, which postulated that a singularity in the framework of gravity coupled to scalars is good if the potential remains bounded from above.  This is intimately related to the so-called ``cloakability'' of the singularity, i.e., the appearance of a continuous family of black hole solutions with a horizon around the singularity at a finite temperature. This horizon criterion is, however, stronger than the mere demand that the potential is bounded from above, and was shown to be in conflict with solutions that are generically considered acceptable, such as the $N=4$ SYM Coulomb branch, as was already noted in \cite{Gubser:2000nd}. One can propose alternative conditions on the admissibility of singularities, motivated by different reasoning. For instance, the computability bound of \cite{Kiritsis:2016kog} requires that the small fluctuation spectrum can be computed by imposing only normalizability as an IR boundary condition.

Violating each of the bounds described above does not automatically have the same physical implications for the solution in question. In fact, some of the bounds are necessary, yet not sufficient, conditions for a singularity to be physically acceptable, while others are believed to be overly strict.  Our main goal in the present paper is to develop a swampland-inspired approach, where the bad singularities are those that truly belong in the swampland, i.e., there is no UV-complete quantum-gravitational theory that would resolve the singularity, either by smoothening it out or by placing it behind a horizon. 

Fortunately, there already exists an extensive framework within the Swampland Program \cite{Vafa:2005ui} treating curvature singularities located at finite spacetime distance but at infinite field distance. This is the concept of \emph{Dynamical Cobordism}, introduced initially \footnote{The term dynamical cobordism was first encountered in \cite{Hellerman:2010dv} to describe solutions interpolating between different string compactifications. Here we refer to a different notion, developed more recently within the Swampland Program. } in \cite{Buratti:2021yia, Buratti:2021fiv} and further developed in  \cite{Angius:2022aeq, Blumenhagen:2022mqw, Angius:2022mgh, Blumenhagen:2023abk, Angius:2023xtu, Huertas:2023syg, Angius:2023uqk, Angius:2024zjv, Huertas:2024mvy, Angius:2024pqk, Altavista:2026edv,Calderon-Infante:2026ymy}, that postulates that if locally, close to the singularity, certain scaling relations hold, the singularity signals the presence of an End-of-the-World (ETW) brane, beyond which the spacetime should end. The existence of such spacetime-ending boundaries is strongly supported by the Cobordism Conjecture \cite{McNamara:2019rup}. 

Let us take a step back, and briefly explain the concept of cobordism \footnote{In this paper, we will use the terms bordism and cobordism interchangeably. Historically, the generalized homology theory was initially referred to as \emph{cobordism}, because the bounding manifolds are the \emph{common border} of the interpolating manifold of one dimension higher. In the mathematical literature, the term \emph{bordism} is used for the generalized homology, and \emph{cobordism} is reserved for the cohomological theory, whereas in the physics literature, the term \emph{cobordism} is more common. In this paper, we will use both terms referring to the homological theory, reserving the use of bordism for the more formal aspects while keeping the term cobordism when discussing swampland-related aspects, in accordance with the naming of the relevant conjectures.}. Motivated by the ubiquitous topology-changing procedures in theories of quantum gravity, a cobordism group $\Omega_k^\xi$ describes equivalence classes of $k$-dimensional smooth compact manifolds, up to topology-changing transformations that preserve a certain physical structure $\xi$. A widely relevant structure for string theories is the Spin structure, which demands that the manifolds in question admit (neutral) fermions. These equivalence classes form a group under the disjoint union of manifolds. Within the framework of generalized symmetries, see e.g., \cite{Gaiotto:2014kfa,Schafer-Nameki:2023jdn}, a non-vanishing cobordism group $\Omega_k^{\xi}\neq 0 $ for a $d$-dimensional theory leads to a global $(d-k-1)$-form symmetry. In view of the conjectured absence of global symmetries in quantum gravity \cite{Banks:2010zn}, it was proposed in \cite{McNamara:2019rup} that the bordism groups of quantum gravity $\Omega_{k}^{QG}$ should all vanish, for $k\leq d$. The key point here is the \emph{quantum gravity structure}, which is currently unknown. One hence has to work iteratively, starting with some reasonable approximate structure $\widetilde{QG}$ such that $\Omega_k^{\widetilde{QG}}\neq 0$ for some $k$, and then figure out what modification of the structure can trivialize the associated global symmetry, either by breaking or gauging it. This has been proven to be a particularly fruitful avenue of pre- and post-dicting objects in string theory, such as in \cite{Dierigl:2022reg, Dierigl:2023jdp, Hamada:2025duq, Heckman:2025wqd, Nevoa:2025xiq, Cavusoglu:2026xiv}.  Several of these objects are in fact non-supersymmetric, and their stability can be intimately related to their bordism charges, such as in \cite{Kaidi:2023tqo, Kaidi:2024cbx, Heckman:2025wqd}. It is worth noting that a major advantage of the cobordism approach is that it is not at all reliant on supersymmetry. However, the purely formal approach of computing bordism groups for increasingly intricate structures can become technically demanding; see, for instance, \cite{Blumenhagen:2022bvh, Basile:2023knk, Debray:2023yrs, Kneissl:2024zox, Braeger:2025kra, Chakrabhavi:2025bfi}.

The Dynamical Cobordism proposal, on the other hand, allows us to reach useful conclusions about a solution simply by computing the scaling of the scalar curvature locally, close to a singularity. The solution can be locally characterized by a single parameter, the critical exponent $\delta$, appearing as $R\sim e^{\delta D}$, with $D$ the field space distance. However, in its present formulation, it lacks predictive power. 
In the spirit of the Swampland Program, and in particular given the similarity to the Swampland Distance Conjecture \cite{Ooguri:2006in} that postulates that for infinite distances $D \to \infty$ in a moduli space $\mathcal{M}$ there arise infinite towers of particles with mass $m \sim e^{-\lambda_{SDC} D}$, one could envision that the specific value of $\delta$ can be constrained \footnote{Note that the field space distance $D$ is only a moduli space distance in the case of a vanishing potential, hence the Dynamical Cobordism statement has broader applicability compared to the Distance Conjecture.}. The Sharpened Distance Conjecture \cite{Etheredge:2022opl} postulates that $\lambda_{SDC} \geq \sqrt{ \frac{1}{d-2}}$. The value of $\lambda_{SDC}$ is not only constrained, but in combination with the Emergent String Conjecture \cite{wald1997} provides information about the physical nature of the tower in question, with only $\lambda_{SDC}= \sqrt{\frac{1}{d-2}}$ being compatible with an Emergent String tower.

In the present paper, we propose a sharpened version of Dynamical Cobordism, which can be interpreted as a quantitative criterion for the compatibility of certain solutions with a UV-completion. While one could envision that a strict bound on $\delta$ is in order \footnote{In fact, bounds on $\delta$ have been proposed in different contexts in \cite{Angius:2022aeq, Blumenhagen:2023abk}, or more recently in \cite{Calderon-Infante:2026ymy}, see note below.}, we find that a certain kind of iterative procedure is in order, similar to that applicable to the original Cobordism Conjecture.

There, one can simply look at a bordism group, and depending on whether it is trivial\footnote{A non-trivial bordism group can also lead to a non-problematic EFT, as long as one compactifies on a manifold belonging to the trivial class. This is referred to as ``gauging'' in \cite{McNamara:2019rup}, in the sense that the bordism class over the compact manifold is zero. }, decide whether a modification is needed to make the theory ``UV-completable''. However, whenever one identifies a singular solution as a dynamical cobordism via the signature scaling relations, there is no concrete way to know whether this necessitates the introduction of a new cobordism defect, that would correspondingly modify the cobordism structure of the theory, or whether the defect is already present within the theory. One can try to identify the defect with objects known in string theory, such as Dp-branes, as was done, for instance, in \cite{Buratti:2021fiv, Angius:2022aeq}, or try to identify novel solutions of the equations of motion that locally generate the same singularity scalings, in the spirit of \cite{Blumenhagen:2022mqw,Blumenhagen:2023abk}.

We argue that, according to the specific value of the critical exponent $\delta$, one can deduce whether the singular dynamical cobordism solution can be UV-resolved within the same theory, or whether, in contrast, a major alteration, such as  a new type of defect or some gauge degrees of freedom, is required.  This can be viewed as a concrete test on whether the ``cobordism structure'' detected by the EFT is compatible with quantum gravity, at least for the solution of interest, or whether a modification is needed\footnote{Note that in principle the cobordism structure might not be fully determined via the field content coupling to gravity in the action, since features such as dualities might be hidden at the level of the action. For the purposes of this work, differentiating between ``structure'' and ``field content'' makes no practical difference, so we will use the two terms interchangeably.}. In a sense, we are now putting Dynamical Cobordism on a similar footing as the original Cobordism Conjecture in terms of predictability. In principle, this strategy can then be applied to different setups that exhibit dynamical cobordism scalings, such as the flux-backtracked solutions of \cite{Apers:2025pon}, that are intimately related with deeper open issues like scale separation in string theory (see \cite{Coudarchet:2023mfs}
for a related review). Hence, Sharpened Dynamical Cobordism would be an additional tool for probing the viability of such setups.

The determination of the allowed region $\mathcal{R}^\xi$ is highly motivated by the singularity criteria outlined above, and we observe that including more fields into the description, i.e., approximating more accurately the true quantum gravity structure, leads to progressively looser constraints for the allowed values of $\delta$. An initially ``bad'' solution can hence become ``good'' upon taking into account the additional structure, but the opposite does not seem to occur, at least for our preferred method of determining the ``good'' regions. To put it differently,  we find that more and more singularities now correspond to the End-of-the-World objects, which is in accordance with the expectation that the modification of the structure leads to the trivialization of more and more bordism groups.

The rest of the paper is structured as follows: With the aim of keeping everything self-contained, section \ref{sec:AllLiterature} is dedicated to reviewing the relevant literature. In section \ref{sec:DynamicalCobordismDefinition}, we review dynamical cobordisms and introduce the defining critical exponent, while in section \ref{sec:LiteratureCriteria}, we briefly summarize certain criteria for the admissibility of a naked singularity. In section \ref{sec:boundsDC}, we translate the literature criteria to statements on the admissible values of the critical exponent. In section \ref{sec:Sharpening}, we formulate our main proposal for a sharpened version of the Dynamical Cobordism Conjecture. We will propose a systematic way to specify the allowed range $\mathcal{R}$ for $\delta$ given a certain action, and work out in detail how this range gets modified upon the introduction of a $q$-form field in the action, which we interpret as a modification of the cobordism structure.
In section \ref{sec:examples}, we will discuss examples violating and respecting the bounds, explaining how this supports our sharpened version of the Dynamical Cobordism Conjecture, and we will close this paper with some concluding thoughts and future directions in section \ref{sec:conclusions}.
We provide some supplementary calculations for the allowed ranges for the critical exponent in the appendices \ref{app:DimensionalReduction}, \ref{app:TopForm} and \ref{app:finite_action}. \linebreak

\emph{
\textbf{Note}: During the later stages of this work, the paper \cite{Calderon-Infante:2026ymy} appeared, overlapping with certain parts of our work. Our work was completed independently, and parts of it had already been presented in past events, such as Swamplandia 2025 and the 2026 CERN Winter School. Throughout the present paper, we will comment and illustrate the similarities and differences between the two works when appropriate.
}

\section{Dynamical Cobordism and Gravitational Bounds}
\label{sec:AllLiterature}

The following section summarizes the relevant parts of the literature that we will use to motivate and formulate the Sharpened Dynamical Cobordism Conjecture. The reader familiar with these topics may skip ahead to section \ref{sec:Sharpening}, where we discuss novel results.

\label{sec:DynamicalCobordism}

\subsection{Dynamical Cobordism and transitions to nothing}
\label{sec:DynamicalCobordismDefinition}

\begin{figure}[!htbp]
    \centering

    \begin{subfigure}[t]{0.48\textwidth}
        \centering

        \begin{minipage}[c]{0.48\linewidth}
            \centering
            \includegraphics[width=\linewidth,height=3.2cm,keepaspectratio]{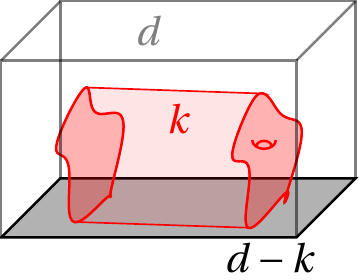}
        \end{minipage}
        \hfill
        \begin{minipage}[c]{0.48\linewidth}
            \centering
            
            \vspace{0.5cm}            \includegraphics[width=\linewidth,height=3.6cm,keepaspectratio]{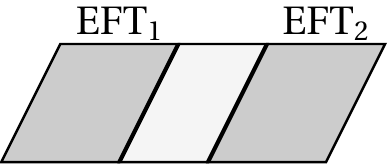}
        \end{minipage}
        \caption{}
        \label{fig:panel_cobordant}
    \end{subfigure}
    \hfill
    \begin{subfigure}[t]{0.48\textwidth}
        \centering
        \begin{minipage}[c]{0.48\linewidth}
            \centering
            \includegraphics[width=\linewidth,height=3.2cm,keepaspectratio]{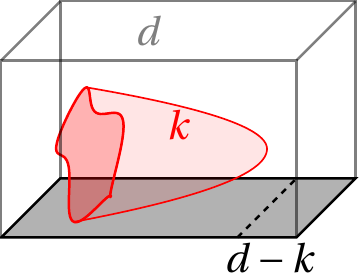}
        \end{minipage}
        \hfill
        \begin{minipage}[c]{0.48\linewidth}
            \centering
            \includegraphics[width=\linewidth,height=3.2cm,keepaspectratio]{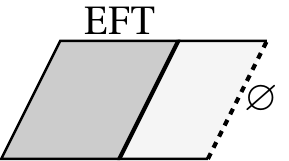}
        \end{minipage}
        \caption{}
    \label{fig:panel_nothing}    
    \end{subfigure}

    \caption{Pictorial representation of cobordant configurations with corresponding EFTs. Side~\subref{fig:panel_cobordant} depicts $k$-dimensional manifolds embedded in a $d$-dimensional spacetime, the  corresponding $(d-k)$-dimensional EFTs  and the interpolating Domain Wall. Side~\subref{fig:panel_nothing} shows the case of a $k$-dimensional manifold cobordant to nothing, together with the corresponding codimension-1 end-of-the-world configuration (ETW domain wall) in the compactified $(d-k)$-dimensional theory.
    }
    \label{fig:cobordism_multipanel}
\end{figure}

Let us start this section with a quick review of Dynamical Cobordism. As already mentioned, the idea stems from the Cobordism Conjecture, which postulates that all the bordism groups in a theory of quantum gravity, $\Omega_{k}^{\rm QG}$, must be trivial to comply with the required lack of global symmetries. In practice, this is often applied to the $k$-dimensional manifold on which a $d$-dimensional theory is compactified. A transition between two cobordant $k$-manifolds then corresponds to a $(d-k-1)$-dimensional object from the $d$-dimensional point of view, or a codimension-1 domain wall within the $(d-k)$-dimensional EFT, conventionally called ``cobordism defect''. When a bordism class is trivial, all the manifolds in it are bordant to nothing, i.e., they are themselves the boundary of a $(k+1)$-manifold with the same structure. From the EFT point of view, now the domain wall is interpolating between the $(d-k)$-dimensional EFT and ``nothing'', i.e., the spacetime ceases to exist past the cobordism defect. This type of object is conventionally called ``End-of-the-World'' (ETW) brane, a term initially introduced in a holographic context for instance in \cite{Raamsdonk:2020tin, VanRaamsdonk:2021duo}. Today ETW branes are considered in many contexts, including in relation to bubbles of nothing/something \cite{Muntz:2025ltu, Hassfeld:2023kpu}, although originally such bubbles were considered as instabilities of the vacuum \cite{Witten:1981gj}. We provide a sketch of cobordant configurations and the corresponding EFTs in figure \ref{fig:cobordism_multipanel}.

Dynamical Cobordism  provides a dynamical realization for these transitions to nothing, via identifying a spacetime singularity with an ETW brane as long as the following three conditions hold \cite{Buratti:2021yia, Buratti:2021fiv, Angius:2022aeq}: 

\begin{itemize}
    \item The singularity is located at finite spacetime distance $\Delta$ from a chosen point of reference.
    \item The field distance $D$ from the point of reference to the singularity goes to infinity. Note that here we do not necessarily impose that this is a moduli space, i.e., one cares about the distance the field traverses regardless of the potential. This is different both from the Distance Conjecture \cite{Ooguri:2006in} that assumes a strict moduli space, or more recent variations of the Distance Conjecture that try to appropriately take into account the potential in the measured distance, such as \cite{Basile:2023rvm, Debusschere:2024rmi, Mohseni:2024njl, Demulder:2024glx}.
    \item The following scaling relations hold 
    locally close to the singularity:
\begin{equation}
\label{eq:ScalingRelations}
    \Delta \sim e^{-\delta/2 \, D}, \qquad\qquad R \sim e^{\delta D} \, ,
\end{equation}
where $R$ is the Ricci scalar.  The dynamical cobordism solution is characterized then by the \emph{critical exponent $\delta$}. 
\end{itemize}
Note that this is not the exact definition of dynamical cobordism introduced in the first dynamical cobordism references, where a lot of emphasis was given on the behavior of a possible tadpole \cite{Buratti:2021yia, Buratti:2021fiv}. The notion of the critical exponent was, in fact, only introduced in \cite{Angius:2022aeq}. We, however, believe that the definition above captures the spirit of the original idea while remaining as general as possible; hence, we will use this as our working definition of a dynamical cobordism to nothing, similarly to \cite{Blumenhagen:2022mqw, Blumenhagen:2023abk}. We also want to remark that, as explored in \cite{Buratti:2021fiv}, the concept of Dynamical Cobordism also applies to transitions where the field does not reach an infinite distance at the singularity: there, the cobordism defect is interpreted as an interpolating domain wall between different EFTs, rather than a transition to nothing. We will not consider such transitions in the present work.

In \cite{Angius:2022aeq}, a universal local description for dynamical cobordisms was developed. In the context of a gravity-scalar theory with the action
\begin{equation}\label{eq:ActionScalarField}
    S \sim \int d^{d}x \sqrt{g} \left( R - \frac{1}{2} (\partial \phi )^2 - V(\phi)\right).
\end{equation} 

The following Ansatz realizes the dynamical cobordism in the $z$-coordinate, and one can place a codimension-1 ETW brane at $z=0$, where $\sigma \rightarrow +\infty$:
\begin{equation}\label{eq:ETWBrane_StdAnsatz}
    ds^2_d = e^{-2\sigma(z)} ds^2_{d-1} + dz^2, \qquad \phi=\phi(z).
\end{equation}

Since one is only interested in a local solution very close to the singularity, the particular geometry of the $d-1$ direction is not of much interest, so for simplicity we consider $ds^2_{d-1}$ to be the Minkowski metric. The analysis for non-flat $ds_{d-1}^2$ of constant curvature was performed in the appendix of \cite{Angius:2022aeq}, yielding very similar results. Here, the equations of motion for the scalar field 
and the $zz$ and transversal components of the Einstein equations are
\begin{equation}\label{eq:EOMs_EDTheory}
\begin{aligned}
    & \phi'' - (d-1)\sigma' \phi' - \partial_\phi V=0, \\
    & (d-1)(d-2) \sigma'^2 + V - \frac{1}{2}\phi'^2=0, \\
    & (d-2)\sigma''- \frac{1}{2}\phi'^2=0.
\end{aligned}
\end{equation}
In \cite{Angius:2022aeq}, it was shown that generically the potential takes the exponential form
\begin{equation}\label{eq:ExponentialPotential}
    V = C_V e^{\delta_V \phi},
\end{equation}
where $C_V$ and $\delta_V$ are constants. Close to the singularity, we can recover the scaling relations \ref{eq:ScalingRelations} with 
\begin{equation}\label{eq:CriticalExponent_GeneralA}
\delta =2\sqrt{\frac{d-1}{d-2}(1-a)}\, ,
\end{equation}
where $a=\frac{V}{V-T}<1$, with $V$ the potential and $T$ the kinetic energy of the scalar. The value of $\delta$ for $a=0$, i.e.  for negligible potential, $\delta_{0}=2\sqrt{\frac{d-1}{d-2}}$, is of special importance, since its relation to $\delta_V$ of the potential directly determines whether one has a scaling solution or kination. We refer the reader to \cite{Calderon-Infante:2026ymy} for a systematic classification.

Multiple examples in the literature have been shown to correspond to dynamical cobordism solutions, some of them after appropriate compactifications that bring the solution in the form \eqref{eq:ETWBrane_StdAnsatz}. A systematic determination of the corresponding $\delta$ values was performed in \cite{Angius:2022aeq}, and among the most interesting cases one finds $p\neq 3$ $Dp$-brane solutions with $p$-dependent $\delta$'s, massive IIA string theory \cite{Romans:1985tz, Bergshoeff:1996ui} with $\delta=5/\sqrt{2}$, $USp(32)$ strings \cite{Sugimoto:1999tx, Dudas:2000ff} with $\delta = 3/\sqrt{2}$, EFT strings \cite{Lanza:2020qmt, Lanza:2021udy} with $\delta = 2\sqrt{2}$, the $D2/D6$ system on a $T^4 \times S^2$ with $\delta=\sqrt{2}$, the Klebanov-Strassler solution \cite{Klebanov:2000nc, Klebanov:2000hb}\footnote{In fact, \cite{Angius:2022aeq} claims the KS solution corresponds to a critical exponent $\delta = 2\sqrt{30}/3$. In \cite{Calderon-Infante:2026ymy} it was recently clarified that the result of \cite{Angius:2022aeq} corresponds to Klebanov-Tseytlin solution, and the Klebanov-Strassler one actually has $\delta= \sqrt{30}/3$.}  and Witten's bubble of nothing \cite{Witten:1981gj}.

\subsection{Criteria for naked singularities}
    \label{sec:LiteratureCriteria}

\label{sec:Bounds}

In this section, we briefly highlight some approaches that have been used to investigate the physical interpretation of a naked singularity in a gravitational theory. 
Note that we do not aim at presenting an exhaustive list of \textit{every} criterion of the literature, but we have rather focused on the Gubser \cite{Gubser:2000nd} and computability \cite{Kiritsis:2016kog} bounds, as those are directly applicable on EFTs. For instance, we will not discuss the Maldacena-Núñez bound \cite{Maldacena:2000mw}, since its regime of applicability differs and requires the full uncompactified theory, making direct comparisons with the EFT results impractical. 

\subsubsection{Gubser criteria}
\label{sec:GubserCriteria}

These criteria \cite{Gubser:2000nd} were formulated in the context of AdS/CFT, for bulk geometries in five dimensions that satisfy the equations of motion of a theory containing gravity and a fixed number of scalar fields. More precisely, they were phrased for codimension-1 geometries of the form of \eqref{eq:ETWBrane_StdAnsatz} with $AdS_5$ asymptotics that reach some singularity deep in the bulk where $\sigma \rightarrow +\infty$. While the initial formulation was in AdS/CFT, it is not unreasonable to consider these bounds in a broader context as well. This was done very recently, for instance, in \cite{Calderon-Infante:2026ymy}, and we adopt their naming conventions here for clarity.

\subsubsection*{Horizon Criterion}

The Gubser \emph{horizon criterion}~\cite{Gubser:2000nd} \footnote{Note this is different than the horizon criterion of \cite{Hassfeld:2025hjx} that classifies spacetimes as problematic in the presence of a cosmological horizon.} establishes that for a naked singularity to be physical,  there must exist a family of finite-temperature $T$ black hole solutions covering it, such that the original singularity is obtainable at the $T\rightarrow 0$ limit. In a holographic context, the Hawking temperature of the black hole serves as an effective infrared cut-off for the dual field theory. If the singularity were not covered by the horizon in the $T\neq 0$ case, then the dual theory could have pathologies related to the naked singularity. 
One possible way of interpreting this condition is as follows: Consider a family of black hole solutions parametrized by the location of the horizon $z=z_*$. The limiting case $z_* \rightarrow 0$ should correspond to a naked singularity. The criterion of Gubser requires that a generic naked singularity should be obtained in this way, providing hence a necessary, yet not sufficient, condition for its admissibility.

Applying this criterion in general requires, at least, specifying the asymptotics of the black hole solutions we want to use to cloak off the singularity. 
A (weaker) local condition can be found such that naked singularities violating it cannot be obtained as the $T \rightarrow0$ limit of a family of black hole solutions, and it will be of interest to us. Even though Gubser's original work was for $d=5$, we now briefly present this condition for general $d$. 

The first step is to introduce a blackening function $h(z)$ in the geometry
\begin{equation}\label{eq:NakedCoveredSingularity}
    ds^2 = e^{-2\sigma(z)} (-h(z)dt^2 + ds^2_{d-2}) + \frac{dz^2}{h(z)}.
\end{equation}
Next, we assume the presence of a horizon at $h(z_*)=0$. The functions $\sigma(z)$ and $h(z)$ are determined by solving the equations of motion for the specific theory under consideration. 

The $zz$-component of the Einstein tensor depends only on the geometry \eqref{eq:NakedCoveredSingularity} and is

\begin{equation}
G_{zz} = \frac{d-2}{2h}  \left( h \sigma'^2 (d-1) - \sigma' h'\right) \quad \implies \quad h(z_*)G_{zz}(z_*) = -\frac{d-2}{2} \sigma'(z_*) h'(z_*), 
\end{equation}
where the evaluation is performed at the horizon. As pointed out in \cite{Gubser:2000nd}, to ensure a positive mass black hole, $h'(z_*)>0$ must hold, while $\sigma'(z_*) <0$ is obtained by examining the other equations of motion. Consequently, $h(z_*)G_{zz}(z_*)\ge0$, which in turn leads to a condition on $T_{zz}$:
\begin{equation}\label{eq:GubserCorolary}
    h(z_*) T_{zz}(z_*) \ge 0, \qquad T_{\mu \nu} = \frac{-2}{\sqrt{g}} \frac{\delta S}{\delta g^{\mu \nu}}.
\end{equation}

In an Einstein-Dilaton (ED) theory, \eqref{eq:GubserCorolary} becomes
\begin{equation}\label{eq:GubserNegativePotential}
    V(z_*) \le 0.
\end{equation}
Note that this bound comes from a local analysis of the EOMs. Using appropriate boundary conditions, the bound can become more stringent. For instance, for AdS asymptotics with cosmological constant $\Lambda \sim V_{AdS}$, it becomes \cite{Gubser:2000nd}
\begin{equation}\label{eq:GubserHolographicCondition}
    V(z_*) \le V_{AdS} <0.
\end{equation}
In the $T \rightarrow0$ limit, the horizon goes to zero, and we obtain a naked singularity. Thus, the previous condition is applicable in the vicinity of the singularity, and one can demand
\begin{equation}\label{eq:NegativePotentialCondition}
    V \le 0 \qquad \text{when approaching the singularity.}
\end{equation}
For example, $V \rightarrow - \infty$ is allowed, but $V \rightarrow + \infty$ is not. Although the criterion 
was originally formulated in holographic settings, it is based on a physical argument that can be reasonably applied to broader settings.

We also want to emphasize that the local bound \eqref{eq:NegativePotentialCondition} on the potential is a necessary, yet not sufficient, condition for the development of a horizon around the singularity.

It is significant to mention that the Gubser horizon criterion has been studied in the framework of singular Renormalization Group (RG) flows in ED theories in \cite{Kiritsis:2016kog, Kiritsis:2025ytb}. The bulk geometry is described by \eqref{eq:ETWBrane_StdAnsatz} and the RG flows are embedded into an ED theory. 
The singularity is located at $\sigma \rightarrow +\infty$, the IR from the dual field theory point of view. In this context, the criterion results in a bound on the growth of the potential when approaching the singularity in one of the possible flows to infinity. 
From an action of the form \eqref{eq:ActionScalarField} with a potential $V \sim-e^{b\phi}$, one finds 
\begin{equation}\label{eq:1USE_NOTGubserBound}
     b < \sqrt{2}\sqrt{\frac{d-1}{d-2}}.
\end{equation}
This is sometimes simply referred to as the Gubser bound \cite{Kiritsis:2016kog, Kiritsis:2025ytb}. 
In \cite{Gouteraux:2011ce}, the Gubser horizon criterion was examined for a black hole solution with AdS asymptotics in an Einstein-Dilaton-Maxwell theory, resulting in a condition relating the exponent of the potential $b$ and the gauge coupling exponent. If the action is 
\begin{equation}
    S \sim \int d^dx \sqrt{g} \left( R - \frac{1}{2}(\partial \phi)^2 -V(\phi) -\frac{1}{2 q!}e^{a\phi} F_q^2\right),
\end{equation}
then one gets the bound
\begin{equation}\label{eq:GaugeFieldAdditionalConstraints}
    -b - a b +2 >0, \qquad 2(d-2) + (d-1)a^2 + 2a b- (d-3)b^2 >0,
\end{equation}
which implies that the values of $b$ and $a$ cannot be completely arbitrary. 

\subsection*{Potential Criterion}

The horizon criterion is known to be too restrictive in some settings. For example, already in \cite{Gubser:2000nd} it was argued how certain states of the Coulomb branch of $\mathcal{N}=4$ SYM theory violate the criterion, but there is no other reason to consider the solutions unphysical. More recently, it was shown in \cite{Calderon-Infante:2026ymy} that certain moduli space flows also violate this condition.

This serves as a motivation 
to introduce the Gubser \textit{potential criterion}~\cite{Gubser:2000nd}, according to which
\begin{equation}
    \text{$V(\phi)< +\infty $ when approaching a singularity.}
\end{equation}
This is weaker than the horizon criterion just by comparison with \eqref{eq:GubserNegativePotential}, and it actually is compatible with the previously mentioned states of the 4d $\mathcal{N}=4$ SYM theory or the moduli space singular flows of \cite{Calderon-Infante:2026ymy}. 
Note that this criterion was postulated to be necessary for a singularity to be good, but it is not sufficient, as examples violating the Breitenlohner-Freedman bound~\cite{Breitenlohner:1982jf} can actually pass the potential criterion.

\subsubsection{Computability criterion}

\cite{Kiritsis:2016kog} addressed the physical meaning of singularities in the renormalization group flows in holographic ED theories. In particular, they studied the RG flow to the boundary of field space ($\phi  \rightarrow + \infty$) corresponding to a dual field theory IR singularity. 
Close to this singularity, the bulk geometry is of the form \eqref{eq:ETWBrane_StdAnsatz}, with $\sigma \rightarrow +\infty$.

The computability criterion demands that the fluctuation problem around any solution must be well-posed without introducing IR boundary conditions except for normalizability \cite{Kiritsis:2016kog}, i.e., the UV data of the boundary are enough to uniquely determine the evolution of the perturbations.

We note that this was not claimed to be a \emph{necessary} condition for the ``goodness'' of the singularity, since in the case it is violated, an embedding of the solution to a more complete framework might still show it is acceptable.

Let us try to give a more quantitative description of the bound. The bulk geometry solves the equations of motion of a $d$-dimensional ED theory with potential $V(\phi)$. The computability criterion imposes a bound on the growth of the auxiliary scalar function called fake superpotential $W(\phi)$ \cite{Kiritsis:2016kog}, defined via 
\begin{equation}
    V = \frac{1}{2} \left( \frac{d W}{d\phi}\right)^2- \frac{d-1}{2(d-2)}W^2.
\end{equation}

This can be translated to a bound on the profile of $\sigma$ corresponding to the RG flow
\begin{equation}
    \sigma(z)=\frac{-2}{(d-2)\gamma^2} \ln(z), \qquad \gamma <\sqrt{2}\sqrt{\frac{d+1}{3(d-2)}}.
\end{equation}

A systematic classification of flows was performed in \cite{Kiritsis:2016kog}, where it was found that only potentials with 
\begin{equation}
\label{eq:computability_pot}
    V \sim-e^{\gamma\phi}, \qquad \gamma < \sqrt{2}\sqrt{\frac{d+1}{3(d-2)}},
\end{equation} 
satisfy the computability criterion. Note that $\gamma=b$ corresponds to our conventions in \eqref{eq:ActionScalarField}. This bound is always stronger than \eqref{eq:1USE_NOTGubserBound}.

\subsection{Interplay with Dynamical Cobordisms}
\label{sec:boundsDC}
A question that arises naturally is whether the criteria mentioned above can be directly translated into the dynamical cobordism language. Using the bounds on the exponent of the potential $b$ in the case of Gubser and computability bounds, we can restrict the critical exponent $\delta$. We want to emphasize that the dynamical cobordism framework is much broader than the one in which the criteria were initially formulated, hence this ``translation'' is a good-faith generalization to more general setups. Note that a similar good-faith extension of applicability of the Gubser bound was considered in \cite{Calderon-Infante:2026ymy}. They recently proposed a bound for the critical exponent, arising as a \emph{geometrization} of Gubser's potential criterion:
\begin{equation}
    \delta \le 2 \sqrt{\frac{d-1}{d-2}}.
\end{equation}
This is thought to be a necessary condition for a $\textit{good}$ singularity. This is weaker than the original Gubser bound, since now positive divergent potentials are allowed, as long as they satisfy $|V|\lesssim \exp\big( \sqrt{2\frac{d-1}{d-2}} \phi\big) $.

Before presenting the remaining literature criteria in the dynamical cobordism language, we want to clarify that, although one can often read off the critical exponent $\delta$ directly from the potential as $\delta=\sqrt{2}b$, this is not true in general\footnote{See also \cite{Calderon-Infante:2026ymy} for a related discussion and classification of cases.}. This may only happen in cases where the potential grows at a rate comparable to the kinetic term, and for the solutions where this is true, the Gubser and computability bounds \eqref{eq:1USE_NOTGubserBound} and \eqref{eq:computability_pot} are in one-to-one correspondence with bounds on $\delta$. We will call this type of solution \emph{dominant}. However, there is a second class of solutions, where now the critical exponent is not determined by the potential, which may be negligible, but instead takes the value $\delta_0=2 \sqrt{\frac{d-1}{d-2}}$. In that case, we need to keep track of the growth of the potential in addition to bounding $\delta$. We will call this type of solution \emph{subdominant}. Note that, in this case, the computability bound is automatically violated.

We summarize all the singularity bounds translated in the dynamical cobordism language
in table \ref{tab:CriteriaDCInterplay}. By the Gubser horizon criterion, we are referring to the necessary condition \eqref{eq:GubserCorolary}, as we cannot know the asymptotics of the singularity described by \eqref{eq:ETWBrane_StdAnsatz}.  
 
\begin{table}[!htbp]
\centering
\begin{tabular}{c| c| c}
\toprule
          Criterion     & Dominant ($\delta=\sqrt{2}b$)           & Subdominant ($b<\delta_0/\sqrt{2}$)                                 \\ 
          \midrule
Gubser horizon~\cite{Gubser:2000nd, Kiritsis:2016kog} & $\delta<\delta_0$                              & $\delta=\delta_0$ and $V\le0$                                                 \\ \hline
Gubser potential~\cite{Gubser:2000nd} & $\delta<\delta_0$                              & $\delta=\delta_0$ and $V\le0$                                                 \\ \hline
Computability~\cite{Kiritsis:2016kog}    & $\delta<2\sqrt{\frac{1}{3} + \frac{1}{d-2} }$ & - \\ \hline
Geometrization of Gubser~\cite{Calderon-Infante:2026ymy}  & $\delta<\delta_0$                              & $\delta=\delta_0$ \\ \bottomrule
\end{tabular}
\caption{ Correspondence between literature criteria of section \ref{sec:LiteratureCriteria} with bounds on critical exponents of dynamical cobordism solutions. \label{tab:CriteriaDCInterplay}}
\end{table}

\section{Sharpened Dynamical Cobordism}
\label{sec:Sharpening}

A lot of progress within the Swampland Program pertains to refining and sharpening conjectures, studying their interrelations with other statements, and improving their predictive power. A prime example is the Swampland Distance Conjecture (SDC) \cite{Ooguri:2006in}, which in its initial formulation claimed that as a modulus $\phi$ traverses an infinite distance $D$ within a moduli space $\mathcal{M}$, an infinite tower of states of mass scale $m$ should become light as $m\sim e^{-\lambda D}$. The nature of the towers was not clear in the initial formulation, but according to the Emergent String Conjecture \cite{Lee:2019wij}, the towers can correspond either to an Emergent String or to a Decompactification limit.

The rate at which this tower becomes massless has been the subject of extended work. The Sharpened Distance Conjecture \cite{Etheredge:2022opl} postulates that there must be at least a tower with $\lambda \geq \frac{1}{\sqrt{d-2}} $, with the saturation of the bound occurring in an Emergent String limit. Other lower bounds have been proposed, for instance, in \cite{Gendler:2020dfp, Lanza:2020qmt, Andriot:2020lea}. In a similar vein, the Convex Hull SDC \cite{Calderon-Infante:2020dhm} imposes a minimum value on the charge-to-mass vectors for the towers. This has been systematically studied across multiple non-trivial examples, e.g., in \cite{Etheredge:2024tok, Etheredge:2023odp}.

As the reader might have already noticed, the Dynamical Cobordism Conjecture bears significant similarities to the SDC. Similarly to how for SDC at the limit of infinite moduli distance the EFT breaks down due to the appearance of the infinite light tower, for dynamical cobordism the infinite field distance limit is accompanied by a curvature singularity. It would be tempting to expect a lower bound in the rate at which the solution becomes singular, i.e., some relation of the form $\delta \geq \delta_{\rm min}$ for the critical exponent. In fact, such a relation was already observed in \cite{Blumenhagen:2023abk}, where it was shown that for two wide classes of examples, a generalized Dudas-Mourad model \cite{Dudas:2000ff} and a generalized Blumenhagen-Font model \cite{Blumenhagen:2000dc}, one has $\delta \geq 2\sqrt{\frac{d-1}{d-2}}$.

However, as we discussed in the previous section \ref{sec:Bounds}, in specific frameworks such as in the case of a scalar-gravity theory with an exponential potential $V\sim -e^{b \phi}$, the criteria for a good singularity can be directly translated into upper bounds for $b$, which, in turn, translates into an upper bound for $\delta$.

The two previous approaches might seem contradictory at first glance. The goal of the present paper is to argue that they are, in fact, compatible, but a more careful look at the interpretation is needed. 

\subsection{Motivation}
Let us, for the time being, consider once again the action 
\begin{equation}
    S \sim \int d^{d}x \sqrt{g} \left( R - \frac{1}{2} (\partial \phi )^2 - C_V e^{b \phi}\right) \, ,
\end{equation} 
with $\phi$ being the dilaton, and consider a solution with a naked singularity at a finite distance.

This is a frequently occurring setup in string theory, describing setups with uncanceled dilaton tadpoles. Depending on the choice of the parameters $\{d,C_V, b\}$ one can describe the Sugimoto model~\cite{Sugimoto:1999tx}, non-supersymmetric $SO(16)\times SO(16)$ string theory~\cite{Alvarez-Gaume:1986ghj, Dixon:1986iz} and massive type IIA theory~\cite{Bergshoeff:1996ui}. In the case of the Sugimoto model (and $SO(16)\times SO(16)$), a famous singular solution is that of Dudas and Mourad \cite{Dudas:2000ff}, which features a spontaneous compactification on an interval, and in \cite{Blumenhagen:2023abk} the setup was called generalized Dudas-Mourad in reference to this solution.
Blumenhagen, Kneissl, and Wang \cite{Blumenhagen:2023abk} systematically analyzed codimension-1 singular solutions that exhibit dynamical cobordism scalings and showed which type of ETW branes should be expected depending on the value of $b$ in the potential. 
We want to direct the attention of the reader to one particular class of solutions, those that arise for $|b|\geq b_{cr}=\sqrt2\sqrt{\frac{d-1}{d-2}}$ \footnote{Note that this differs from the factor in~\cite{Blumenhagen:2023abk} due to a different normalization.}. In that case, it was shown that there exists a codimension-1 solution that satisfies the scaling relations with $\delta=\sqrt2|b|$.
At face value, the action violates the bound \eqref{eq:1USE_NOTGubserBound}, and the solution also violates the bound on $\delta \leq 2\sqrt{\frac{d-1}{d-2}}$ proposed in \cite{Calderon-Infante:2026ymy}. However, the authors of \cite{Blumenhagen:2023abk} managed to explicitly construct a brane-like object sourcing precisely the required type of singularity.

So, how can the two opposing reasonings be reconciled? A crucial step in finding the solution for the ETW brane was to include in the action a $(d-1)$-form gauge potential, the Hodge dual of the 0-form that generates the dilaton tadpole in the original action. The ETW brane turned out to be charged under this field, which plays a crucial role in interpreting the physical situation.

\subsubsection*{Interpretation}

A specific EFT, described by an action, corresponds to a specific cobordism structure $\Omega^\xi$. At least schematically, solutions within this EFT would correspond to bordism equivalence classes within the group $\Omega^\xi$. In the ideal case where one works with the full theory of quantum gravity, then of course there is only a single trivial equivalence class for all solutions, but we assume that we are starting with an EFT that corresponds to an approximate structure $\widetilde{QG}$. For this specific action, if we wish to match a singular solution fulfilling the prerequisites of the Dynamical Cobordism Conjecture, i.e. the singularity is at finite spacetime distance, with the field displacement going to infinity, and the scaling relations \eqref{eq:ScalingRelations} satisfied, with a genuine End-of-the-World object, the critical exponent $\delta$ needs to be in the \emph{allowed region} $\mathcal{R}^\xi$, where the superscript $\xi$ serves to emphasize this is a structure (or action) -dependent range. To ensure that the interpretation as an ETW object is meaningful, we remain agnostic about the applicability of Sharpened Dynamical Cobordism to singularities covered by horizons, and limit it to naked ones. 

In the case of the action \ref{eq:ActionScalarField} above the allowed region is \emph{$0\leq \delta\leq 2\sqrt{{(d-1)}/{(d-2)}}$}. If $\delta \notin \mathcal{R}$, this should be regarded as a signal that the EFT we are considering is incomplete, and the End-of-the-World object should be an additional defect beyond the initial theory, that needs to be included ``by hand''. 

Conceptually, this corresponds directly to how one works with cobordism groups. Upon encountering $\Omega^{\rm \widetilde {QG}}_k\neq 0$, for some sensible $\widetilde{QG}$, one does not discard the manifolds classified by this non-trivial group. Instead, the cobordism structure is appropriately modified, either via gauging or breaking, such that the group is trivialized.
This is precisely what we think is happening in the motivating example above: The ETW object explicitly constructed in \cite{Blumenhagen:2023abk} is charged, i.e., finding the solution necessitates the introduction of an extra field and the subsequent modification of the action, hence the accompanying structure.

In a sense, a solution with a ``bad'' $\delta$ signals that the bordism group for the selected structure is not trivial, i.e., the spacetime does not end as would be expected in the true $\Omega^{QG}=0$ case. The singular behavior then serves as a warning that issues may arise when one attempts to UV-complete the theory, since a corresponding non-trivial global charge would exist. Luckily, the situation is amenable, and the introduction of an external defect can be enough to break the global symmetry. The value of $\delta$ for the new defect may remain the same, and act as a guideline for the construction of the new defect, but now the allowed region $\mathcal{R}^\xi$ has changed, so that $\delta \in \mathcal{R}^{\tilde \xi}$. All in all, the Sharpened Dynamical Cobordism can be stated as

\subsubsection*{Sharpened Dynamical Cobordism Conjecture:}
    \begin{itemize}[-]
        \item 
    A spacetime singularity located at finite spacetime distance but infinite field distance corresponds to an ETW object only if the dynamical cobordism scaling relations hold and the critical exponent $\delta$ is in the structure-dependent allowed region $\mathcal{R}^{\xi}$. 
    \item A solution with $\delta \notin \mathcal{R}^\xi$ can be made compatible with quantum gravity upon an appropriate modification of the structure. This may include the introduction of additional fields in the action, and/or the inclusion of explicit defects, which can serve as ETW objects.
 \end{itemize}
The determination of the allowed region should follow some concrete physical reasoning. In the present paper, we will use a Gubser-motivated condition. We note that the Gubser criteria were initially formulated in a holographic framework, but here we propose an extension and apply the condition regardless of the background. Within the Swampland Program, such an approach is common and has yielded many insightful results. For instance, the No Global Symmetries Conjecture \cite{Banks:2010zn} is postulated to hold generally, but it has been partially proven in a holographic context \cite{Harlow:2018tng}. Other Conjectures that are better supported in AdS but have been applied to more generic setups include the AdS Distance Conjecture \cite{Lust:2019zwm}, which, extrapolated to dS, and together with experimental input, led to the Dark Dimension Scenario \cite{Montero:2022prj}.

Note that one can possibly construct multiple singular solutions that satisfy the dynamical cobordism scalings for a given action. These solutions are not all necessarily characterized by the same critical exponent. In fact, some could be in the allowed region, while others could correspond to “unphysical” values of $\delta$. If our interpretation is correct, this is something to be expected. Even a non-trivial bordism group includes a trivial class, and these ''good'' solutions could precisely correspond to this case. 
  
Finally, we want to emphasize that this is a conceptual equivalence, since formally defining and computing bordism groups for singular manifolds is a daunting task. 

\subsection{Specifying the allowed region: Motivating \texorpdfstring{$\mathcal{R}$}{R} a la Gubser}

For our proposal above to be practically useful, one needs a systematic way to determine the allowed region $\mathcal{R}$. This is where we draw inspiration from the bounds on singularities we reviewed in section \ref{sec:GubserCriteria}. 
We will begin by considering an ED theory, where the dynamical cobordism solutions were originally embedded, as explained in section \ref{sec:DynamicalCobordism}. 
Our starting point is the Gubser horizon criterion. This necessitates $V\le0$, which for a dynamical cobordism solution translates to
\begin{equation}\label{eq:DynamicalCobordismBound}
    \delta \le 2 \sqrt{\frac{d-1}{d-2}} = 2 \sqrt{1+ \frac{1}{d-2}}=\delta_0.
\end{equation}
This straightforwardly leads to the allowed region 
\begin{equation}
    \mathcal{R} = \left[0,2 \sqrt{\frac{d-1}{d-2}} \right].
\end{equation}

Despite being motivated by Gubser, we note that it is not equivalent to the Gubser criteria. For $\delta = \delta_0$, we know that solutions with $a=0$ allow a positive divergent potential \cite{Angius:2022aeq}, as we saw in section \ref{sec:DynamicalCobordism}, violating \eqref{eq:NegativePotentialCondition} and the Gubser potential criterion. Then, \eqref{eq:DynamicalCobordismBound} is actually weaker than \eqref{eq:NegativePotentialCondition}.
Our allowed region, in this particular case of an Einstein-Dilaton theory, in fact, coincides with the geometrization bound of \cite{Calderon-Infante:2026ymy}, but we will see that $\mathcal{R}$ will deviate from this bound in the case of more complicated theories.

While in this simple Einstein-Dilaton theory one gets the allowed region trivially, this procedure can be generalized to more involved setups,  using equation \eqref{eq:GubserCorolary} to constrain the behavior of $V(\phi)$ for a generic theory.

Before proceeding to calculate how the presence of a higher-form field motivates this Gubser-motivated allowed region, a comment is in order. We believe this is a reasonable approach, but one could imagine that one could impose some different physical condition to delineate the allowed area. In the appendix \ref{app:finite_action}, we observe how similar, yet in certain cases slightly stricter, constraints can arise by imposing that the singular solution contributes finitely to the action.

\subsection{Modification of \texorpdfstring{$\mathcal{R}$}{R}: Adding a \texorpdfstring{$q$}{q}-form field}
\label{sec:HigherForm}

As a proof-of-concept, we want to apply the same reasoning to a different action and determine the new allowed region $\mathcal{R}$ for $\delta$.
For simplicity, we do this by adding a $q$-form field to the action, which will take a non-trivial value in the solution. The action is now of the following form
\begin{equation}\label{eq:Actino_ScalarQForm}
    S \sim \int d^dx \sqrt{g} \left( R - \frac{1}{2}(\partial \phi)^2 -V(\phi) -\frac{1}{2 q!}e^{a\phi} F_q^2\right),
\end{equation}
where $F_q$ is a $q$-form field strength of a $(q-1)$-form gauge potential, $F_q = d A_{q-1}$ and $a$ parametrizes the $\phi$-dependent gauge coupling. For $q=2$, the theory at hand is the Einstein-Dilaton-Maxwell (EDM) theory. In \eqref{eq:Actino_ScalarQForm} we 
allow for $q$ in the range $1<q<d-1$, namely those that cannot be dualized to a scalar or a potential term.\footnote{This case is discussed separately in appendix B.} The equations of motion are
\begin{subequations}
\begin{align}\label{eq:ETWFlux_EOMs}
    R_{\mu \nu} - \frac{1}{2}R g_{\mu \nu} = \frac{1}{2} T_{\mu \nu}&, \qquad \nabla^2 \phi = \frac{a}{2q!} e^{a\phi} F^2_q + V'(\phi), \qquad d * e^{a\phi}F_q = 0, \\[1.5mm]
    & T_{\mu \nu}^{\phi} = \partial_{\mu}\phi \partial_{\nu} \phi - \frac{1}{2}g_{\mu\nu} (\partial \phi)^2 - g_{\mu\nu} V(\phi), \\[1.5mm]
    & T_{\mu \nu}^{F_q} = \frac{1}{q!} e^{a\phi} \left( qF_{\mu \alpha_2 ... \alpha_q}{F_{\nu}}^{ \alpha_2 ... \alpha_q} - \frac{1}{2}g_{\mu\nu} F_q^2 \right).
\end{align}    
\end{subequations}

Note that the presence of the higher-form field makes the introduction of another warp factor necessary, so the ansatz for the metric is now
\begin{equation}\label{eq:Generalized_Ansatz}
    ds^2_d = e^{-2\sigma(z)} ds^2_{d-q} + e^{-2\xi(z)} ds^2_{q-1} + dz^2.
\end{equation}

With these assumptions,
for an electrically charged configuration, we find that $F_q$ simplifies to
\begin{equation}
    F_q = -Q e^{-a \phi} e^{(d-q)\sigma -(q-1)\xi}\, dz \wedge \Omega_{q-1},
\end{equation}
which corresponds to a point-like configuration with charge $Q$,  
\begin{equation}
    Q= \frac{1}{V_{d-q}}\int_{\mathcal{M}_{d-q}} e^{a\phi}*F_q.
\end{equation}
We assumed the space $\mathcal{M}_{d-q}$ described by $ds^2_{d-q}$ has volume $V_{d-q}$. In a convenient coordinate system, the form $\Omega_{q-1}$ can be taken as $dt \wedge dx^1 \wedge ... \wedge dx^{q-2}$.

Now, departing from the direct approach of solving the EOMs as in \cite{Angius:2022aeq}, we adopt an educated Ansatz inspired by the original dynamical cobordism solution, valid locally close to the singularity and capturing the leading behavior:
\begin{equation}\label{eq:HigherFormLogarithmicAnsatz}
\begin{aligned}
    \sigma = C_\sigma \ln(z), \qquad \xi = C_\xi \ln(z), \qquad \phi =-C_{\phi} \ln(z) + \ln(M_{\phi}), \qquad V= C_V e^{\delta_V \phi}.
\end{aligned}
\end{equation}
The constant $M_{\phi}$ seems to be irrelevant in the limit $z\rightarrow 0$. However, we will see that it is crucial to recover the results of \cite{Angius:2022aeq} and the related interpretation. 

From here, we observe that we find the scaling relations
\begin{equation}
    R \sim 1/z^2, \qquad \Delta \sim z, \qquad D = \frac{-1}{\sqrt{2}}|C_{\phi}|\ln(z),
\end{equation}
from which we can read off the critical exponent $\delta$
\begin{equation}\label{eq:CriticalExponentRecovering}
    \delta = \frac{2\sqrt{2}}{|C_{\phi}|}.
\end{equation}
When substituting in the equations of motion, the contribution of $V(\phi)$ scales like $z^{-C_{\phi}\delta_V}$, while the kinetic term of the scalar and the curvature scale as $z^{-2}$. We need to differentiate between the potential contributing sub-dominantly versus similarly to the other terms. If we impose the latter, we get
\begin{equation}\label{eq:ConvinientDeltaManipulation}
    C_{\phi} = \frac{2}{\delta_V}, \qquad \qquad \delta = \sqrt{2}|\delta_V|.
\end{equation}
So we can relate the critical exponent and the exponent of the potential. At the computational level, it becomes convenient to take \eqref{eq:ConvinientDeltaManipulation} as a definition of $\delta_V$, holding even when the potential is subdominant and $\delta_V$ does not correspond to the exponent of the potential. To be more concrete, this convenient definition includes two cases
\begin{equation}
    V \sim e^{b \phi}, \qquad \text{with } b=\delta_V \text{ or } b <\delta_V.
\end{equation}
Let us systematically discuss the three cases, depending on whether $e^{a\phi}F^2_q$ and $V(\phi)$ contribute to the EOMs (i.e. with a power of $z^{-2}$) or are subdominant:

\subsubsection*{Case 1: Both~\texorpdfstring{$V(\phi)$}{TEXT} and \texorpdfstring{$e^{a\phi}F^2_q$}{TEXT} contribute}
Imposing that every term contributes as $z^{-2}$ we have
\begin{equation}\label{eq:PotentialAndFormContribute}
    C_\phi = \frac{2}{\delta_V}, \qquad C_\sigma = -\frac{a+\delta _V}{(d-q) \delta _V}.
\end{equation}
$C_{\sigma}$ now corresponds to ensuring a contribution of the higher form. $C_\xi$ is found through the scalar field equation, and $Q^2$ and $C_V$ are computed using the Einstein Equations in terms of $M_{\phi}^{-\delta_V}$. Simplifying, the rest of the solution is
\begin{equation}\label{eq:DCGaugeSolution}
\begin{aligned}
        &C_V = -c \frac{\Xi_1 \Xi_2}{(d-q)^2 \delta _V^2 \left(a+(q-1)\delta _V\right){}^2}, \qquad Q^2 =2c^{-a/\delta_V}  \frac{\Xi_0\Xi_2}{(d-q)^2 \delta _V^2 \left(a+(q-1)\delta _V\right){}^2}, \\
        & C_{\xi} = \frac{a}{(q-1)\delta_V} - \frac{\Xi_2}{(d-q)(q-1)\delta_V(a+(q-1)\delta_V)}, \qquad c = M_\phi^{-\delta_V} >0,
\end{aligned}
\end{equation}  
where we have defined the quantities
\begin{equation}
    \begin{aligned}
        \Xi_0 &= 2(d-q) - (d-2)(a+\delta_V) \delta_V, \qquad \Xi_1= 2(q-1)(d-q) + (d-2)(a+\delta_V)a. \\
        \Xi_2 &= 2(q-1)(d-q)+(d-1)a^2 +(q-1)(2a-(d-q-1)\delta_V) \delta_V.
    \end{aligned}
\end{equation}
Note that $C_V$, thus the potential, is not fixed by $\delta$, because the positive constant $c$ is arbitrary. Nonetheless, its sign is fixed by $\delta$. This feature was observed in \cite{Angius:2022aeq}, and for us it will be a key property of dynamical cobordism. Note that this solution only makes sense if $Q^2 \ge 0$. Therefore, imposing that the potential must be non-positive, $C_V \le 0$, we arrive at two separated regions delineated by the conditions
\begin{subequations}
    \begin{equation}\label{eq:NegativePotentialNormalRegion}
        \mathcal{R}_{r}: \qquad \Xi_0 \ge 0, \qquad  \Xi_2 \ge 0,
    \end{equation}
    \begin{equation}\label{eq:NegativePotentialExoticRegion}
        \mathcal{R}_{b}: \qquad \Xi_1 \le 0.
    \end{equation}
\end{subequations}

\subsubsection*{Case 2:~\texorpdfstring{$V(\phi)$}{TEXT} is subdominant}
As commented on before, now we enforce \eqref{eq:ConvinientDeltaManipulation} as a definition, so $\delta_V$ no longer corresponds to the exponent of the potential. Taking the form to contribute as $z^{-2}$, we find \eqref{eq:PotentialAndFormContribute}. The rest of the equations imply two different solutions
\begin{equation}
    \begin{aligned}
            C_\xi= -\frac{2(d-q-1)}{a\delta_V(d-2)},  \qquad 
            Q^2 = \frac{4c^{-a/\delta_V}}{\delta_V^2}\left(\frac{2 (q-1) (d-q-1)}{a^2 (d-2)}+1\right) >0, \qquad \Xi_1=0,
    \end{aligned}
\end{equation}
or 
\begin{equation}
     C_\xi=\frac{a}{(q-1)\delta_V}, \qquad Q^2=0, \qquad \Xi_2=0.
\end{equation}

Thus, we have found that whenever the higher form dominates the contribution of the scalar potential, the critical exponent must be given by the regions of \eqref{eq:DCGaugeSolution} compatible with $C_V=0$, i.e., $\Xi_1=0$ or $\Xi_2=0$. Note that for $\Xi_2=0$ we obtain $Q^2=0$, exactly as \eqref{eq:DCGaugeSolution}. 

\subsubsection*{Case 3:~\texorpdfstring{$e^{a\phi}F^2_q$}{TEXT} is subdominant}
\label{sec:ScalarDynamicalCobordism}
In this case, the equations reduce to those of an ED theory \eqref{eq:EOMs_EDTheory}, and therefore the solution is the same as in \cite{Angius:2022aeq}. However, to present our results in a uniform manner, we will rewrite it here using the ansatz \eqref{eq:HigherFormLogarithmicAnsatz}. The EOMs result in

\begin{equation}\label{eq:DC_Solution_Scalar}
    C_{\phi} = \frac{2}{\delta_V}, \qquad C_{\sigma}= \frac{-2}{(d-2)\delta_V^2}, \qquad C_V=c \frac{2}{\delta_V^2}\left(1-\frac{d-1}{d-2} \frac{2}{\delta_V^2}\right), \qquad c=M_{\phi}^{-\delta_V}>0.
\end{equation}

The sign change occurs precisely at $\delta_0$.
However, unlike in the ED theory, we still need to impose that the term $e^{a\phi}F^2_q$ is subdominant as a consistency condition. This is
\begin{equation}
    C_\sigma(d-q)+a/\delta_V >-1.
\end{equation}
By substituting \eqref{eq:DC_Solution_Scalar} we found that the region compatible with negative potential is
\begin{equation}\label{eq:GreenRegion}
    \mathcal{R}_{a,g}: \qquad \Xi_0 <0, \qquad \delta_V^2 \le2 \frac{d-1}{d-2}. 
\end{equation}
Here, note that we have $C_\xi=C_\sigma$, so it actually corresponds to a codimension-1 ETW. In addition, if the potential is also subdominant, we find $\delta_V^2=2(d-1)/(d-2)$.

Now that the logic by which the allowed region gets modified has become clear, let us summarize our findings and compare them with other similarly applicable criteria in the following section. 

\subsection{The sharpening with a higher form: the region~\texorpdfstring{$\mathcal{R}$}{R} and analysis}
\label{sec:BoundComparison}

\begin{figure}[!htbp]
\centering
\includegraphics[width=0.7\textwidth]{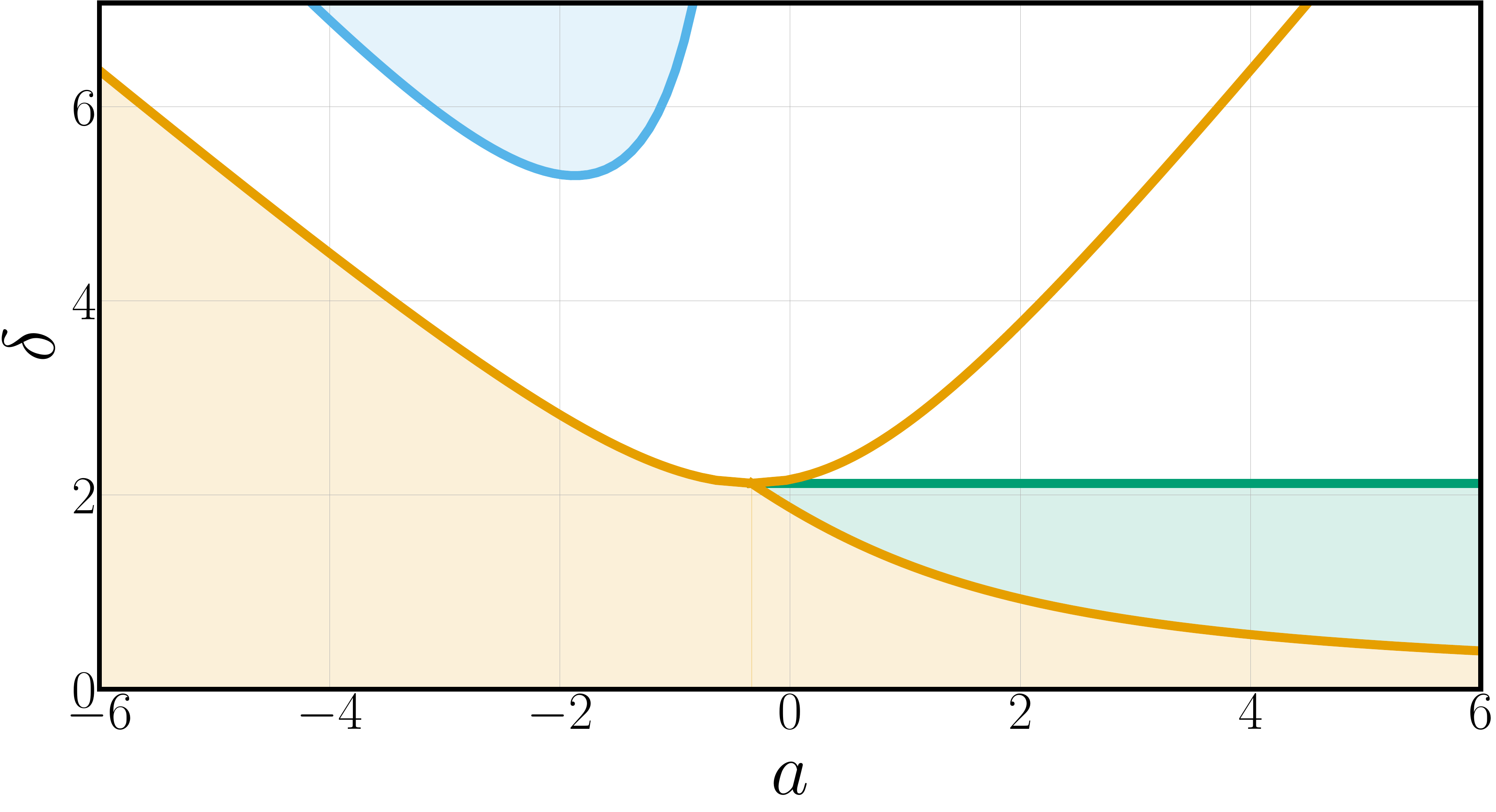}
\caption{Representation of the allowed region in the $(a, \delta)$ plane, adding a three-form and assuming $d=10$. The colors identify the origin of each region. The orange area corresponds to \eqref{eq:NegativePotentialNormalRegion}; the blue to \eqref{eq:NegativePotentialExoticRegion}; and the green to \eqref{eq:GreenRegion}. \label{fig:DeltaAndExponentConstraints}}
\end{figure}

According to the proposed Sharpened Dynamical Cobordism, for a theory with the action \eqref{eq:Actino_ScalarQForm}, the allowed region for the critical exponent $\delta=\sqrt{2} |\delta_V|$ is given by \eqref{eq:NegativePotentialNormalRegion}, \eqref{eq:NegativePotentialExoticRegion} and \eqref{eq:GreenRegion}. For the sake of clarity, we assume $\phi \rightarrow+\infty$ (which at the computational level means $\delta_V >0$). The case with $\phi \rightarrow- \infty$ can be easily worked out in a similar fashion. Then, the regions that form $\mathcal{R}^{s,q}$ are 
\begin{equation}\label{eq:DeltaBoundsHigherForm}
\mathcal{R}:\delta \in 
\begin{cases} 
    [0,\delta_{02}] \cup [\delta_{01}, +\infty) & \qquad \text{if } a < -a_c, \\
    [0, \delta_0] \cup \{\delta_{02}\} \cup[\delta_{01}, +\infty)   & \qquad\text{if } -a_c \le a < 0, \\
    [0,\delta_0] \cup \{\delta_{02}\} & \qquad\text{if } a \ge 0,
\end{cases}
\end{equation}
where
\begin{equation} \label{eq:LimitsHigherForm}
    \begin{aligned}
        &\frac{\delta_{02}}{\sqrt{2}} = \frac{a+\sqrt{(d-q)\left( 2(d-q-1)+\frac{a^2(d-2)}{q-1}\right)}}{d-q-1}, \qquad a_c = \frac{\delta_0}{\sqrt{2}} \frac{q-1}{d-1}, \\[1.5mm]
        &\frac{\delta_{01}}{\sqrt{2}} = - \left( a+\frac{2(q-1)(d-q)}{(d-2)a}\right), \qquad \delta_0= 2 \sqrt{\frac{d-1}{d-2}}.
    \end{aligned}
\end{equation}
These allowed values correspond to the colored regions in figure \ref{fig:DeltaAndExponentConstraints} for $d=10$ and $q=3$. 

To end this section, we remark that the ansatz \eqref{eq:Generalized_Ansatz} is different from the codimension-1 of \cite{Angius:2022aeq} or even the codimension-2 of \cite{Blumenhagen:2022mqw}. However, in appendix \ref{app:DimensionalReduction} we show how, after a compactification, we are able to recover the results described in this section using a codimension-1 ansatz. 

We observe that these bounds are less strict than \eqref{eq:DynamicalCobordismBound}. Thus the addition of another field to the theory allows for more values of $\delta$. This matches our interpretation, because the more structure an EFT has, the more resolutions are possible.

As we have used the Gubser horizon criterion to motivate $\mathcal{R}$, it becomes interesting to examine the exact differences between our proposed bounds and the good singularities of Gubser. We note that fully determining whether the dynamical cobordism solutions can satisfy the Gubser horizon criterion is not possible, since information about the asymptotics is needed to explicitly construct a black hole solution that cloaks the singularity. However, we can use \eqref{eq:NegativePotentialCondition} as a first check to see if our solutions can pass the Gubser criterion.

For an ED theory, \eqref{eq:NegativePotentialCondition} gives $ \delta \le 2 \sqrt{\frac{d-1}{d-2}} = \delta_0 $, 
which is identical to our region $\mathcal{R}$. Up to the equality, this coincides with the Gubser horizon criterion  for the cases where $\delta$ correlates to the exponent of the potential \cite{Kiritsis:2016kog}. The Gubser potential criterion leads to the same result. Similarly, the recently proposed bound of \cite{Calderon-Infante:2026ymy} is in exact agreement with our proposed region.

The situation changes when we include a higher-form field in the action. Since we now have the solution \eqref{eq:DCGaugeSolution} at hand, we will apply \eqref{eq:GubserCorolary} with the solution  to obtain a more precise version of the Gubser bound. By considering electric configurations
\begin{equation}
    F_q = f(z)\, dz \wedge \Omega_{q-1},
\end{equation}
with the geometry \eqref{eq:NakedCoveredSingularity} the condition \eqref{eq:GubserCorolary} is
\begin{equation}\label{eq:1USE_V0MaxwellField}
    V(\phi) + \frac{1}{2} e^{a\phi} e^{2\sigma(q-1)}(f(z))^2 \le 0.
\end{equation}
Note that the only possible way of satisfying the inequality is to have $V <0$, consistently with our methodology. \eqref{eq:1USE_V0MaxwellField} simplifies to
\begin{equation}\label{eq:1USE_ImprovedGubser}
    -\Xi_2\big((d-2)(a+\delta_V)^2+2(d-q)(q-2) \big) \le 0.
\end{equation}
If $\Xi_2 \ge 0$, this inequality is trivially satisfied for every $a$ and $\delta_V$. In figure \ref{fig:DeltaAndExponentConstraints} the orange and green regions satisfy this condition. 
However, the blue region of  figure \ref{fig:DeltaAndExponentConstraints} is situated at $\Xi_2 <0$, so \eqref{eq:1USE_ImprovedGubser} cannot be satisfied. 
This implies that singularities in the blue region of figure \ref{fig:DeltaAndExponentConstraints} cannot be the $T \rightarrow 0$ limit of black holes of the form of \eqref{eq:NakedCoveredSingularity}. On the other hand, the Gubser potential criterion allows for $\delta \ge \delta_{01}$ (the blue region in figure \ref{fig:DeltaAndExponentConstraints}) because $C_V \le 0$.

In the case of $q=2$, i.e., for an Einstein-Dilaton-Maxwell theory, the Gubser horizon criterion for charged black hole solutions with AdS asymptotics was related to the exponents of the scalar field and the gauge coupling \cite{Gouteraux:2011ce}. For the cases where $\delta_V$ is the exponent of the potential, the solutions fulfilled $\Xi_0>0$, $\Xi_2>0$. This is remarkably close to our conditions, which additionally allow for the saturation of the aforementioned inequalities.

Some final comments are in order. In our analysis, we have focused on the leading behavior of the potential. Therefore, a solution with $C_V=0$ does not necessarily imply a vanishing $V$. Any potential is allowed as long as it diverges slower than the growth corresponding to $\delta$ associated with $C_V=0$. This was also observed in \cite{Angius:2022aeq, Calderon-Infante:2026ymy} since solutions with $\delta=\delta_0$ correspond to cases where the kinetic energy of the scalar dominates the potential. Here, we provide the generalization of this observation in the presence of the higher form. A diverging positive potential is now allowed, assuming $\phi \rightarrow +\infty$, for the following combinations of parameters:
\begin{equation}
    \delta = \begin{cases} 
    \delta_{01} & \qquad \text{if } a < 0 \\
    \delta_0    & \qquad \text{if } a > -a_c \\
    \delta_{02} & \qquad \forall a
\end{cases}
\end{equation}

Finally, the difference between the regions characterized by $\Xi_2>0$ and $\Xi_2 \le0$ goes beyond the applicability of the Gubser criteria. Considering the process of appendix \ref{app:DimensionalReduction} of compactifying to a $q$-dimensional theory, the higher-form contribution can be viewed as a potential $V_Q \propto Q^2$, while $V(\phi)$ also contributes as a potential to the lower-dimensional theory $V_q \propto V(\phi)$. The total potential is $V_{eff}=V_Q+V_q$. As  we look for the dominant behavior of the potential, if $V \le0$ is imposed we recover \eqref{eq:NegativePotentialNormalRegion}, \eqref{eq:NegativePotentialExoticRegion} and \eqref{eq:GreenRegion}. However, if we motivate the bound with $V_{eff} \le0$ only \eqref{eq:NegativePotentialNormalRegion} and \eqref{eq:NegativePotentialExoticRegion} (those with $\Xi_2>0$) are included, leaving exponents $\delta \ge \delta_{01}$ (those with $\Xi_2<0$) outside $\mathcal{R}$. We work on the $d$-dimensional theory; thus, the motivation $V\le0$ is more applicable in our picture. All these cases are grouped in table \ref{tab:LimitingDeltas}, where
\begin{equation}
    \delta_{00}= \frac{a}{2} \left( -1 + \sqrt{1+\frac{8}{a^2} \frac{d-q}{d-2}}\right).
\end{equation}

\begin{table}[!htbp]
    \centering
    \begin{tabular}{c|c|ccc}
    \toprule
                  & $d$-dim. potential & \multicolumn{3}{c}{$q$-dim. contributions to potential}   \\       
                  & $V(\phi)$     & \multicolumn{1}{c|}{$V_q \propto V(\phi)$} & \multicolumn{1}{c|}{$V_Q \propto Q^2$} & $V_{eff}= V_q + V_Q$ \\ \toprule
    $\delta =\delta_{00}$ & $-\infty$            & \multicolumn{1}{c|}{$-\infty$}             & \multicolumn{1}{c|}{$\sim 0$}          & $-\infty$            \\ \midrule
    $\delta =\delta_{01}$ & $\sim 0$        & \multicolumn{1}{c|}{$\sim 0$}              & \multicolumn{1}{c|}{$+\infty$}         & $+\infty$            \\ \midrule
    $\delta =\delta_{02}$ & $\sim 0$        & \multicolumn{1}{c|}{$\sim 0$}              & \multicolumn{1}{c|}{$\sim 0$}          & $\sim 0$             \\ \bottomrule
    
    \end{tabular}
    \caption{Behavior of the dominant term of the potential for the different limiting values of $\delta$ in the $d$-dimensional theory or the $q$-dimensional one considered in appendix \ref{app:DimensionalReduction}. The cases $\sim0$ include the subdominant alternatives.\label{tab:LimitingDeltas}}
\end{table}

A physical reason for preferring $V<0$ is the consideration of charged ETW branes in the lower-dimensional theory. Codimension-1 objects can be charged under a top-form, which can later be dualized to a potential. If we just treat its contribution as a potential, the possibility of charged ETWs is ignored. In fact, this situation resembles that of \cite{Blumenhagen:2023abk}, where charged ETW branes were introduced. We will comment more on this in section \ref{sec:examples}.  

To conclude, we have observed how the Sharpened Dynamical Cobordism bounds are a weaker condition than the Gubser potential criterion and, by extension, the computability criterion of \cite{Kiritsis:2016kog}. We will now discuss physical examples where the Sharpened Dynamical Cobordism can be non-trivially tested.

\section{Examples}
\label{sec:examples}

In this section, we want to explore how our sharpened understanding of Dynamical Cobordism fits within concrete setups that feature the required scalings. We will start our discussion by testing the conjecture against some setups that have already been identified as dynamical cobordisms in the literature, i.e., Witten's Bubble of nothing and massive type IIA string theory, and in the latter part of this section we will present some new examples that we have identified as dynamical cobordism solutions, and discuss how their scaling exponents relate to the perceived Cobordism structure.

\subsection{Witten's Bubble of nothing}

The famous Witten's bubble of nothing (BoN) \cite{Witten:1981gj} is an instability that arises when considering a compactification on $\mathbb{M}_4\times S^1$. In particular, the associated instanton corresponds to the $S^1$ shrinking smoothly to zero size over the locus of an $S^3$, and inside this ball bounded by the $S^1$ the spacetime ceases to exist. Wick-rotated to Lorentzian coordinates, this bubble of nothing expands at the speed of light, eventually consuming all spacetime.

In Euclidean coordinates, the gravitational instanton looks like a Schwarzschild black hole in 5d, with the $S^1$ shrinking to zero at the horizon. To get the dynamical cobordism solution, one needs to reduce to 4d along the $S^1$ \cite{Angius:2022aeq}. The 4d solution close to the singularity looks like
\begin{equation}
    ds_4^2 \sim dy^2 +y^{2/3}d\Omega_3^2 \, , 
\end{equation}
and can be identified with a dynamical cobordism solution with critical exponent $\delta =\sqrt{6}$ \cite{Angius:2022aeq}.

Since we are simply dealing with a scalar theory with no potential in 4d, we can straightforwardly determine the allowed region for $\delta$ using \eqref{eq:DynamicalCobordismBound}, namely $\mathcal{R}=[0,2\sqrt{2}]$, and in fact, $\delta \in \mathcal{R}$. This means that the singularity indeed signals an End-of-the-World transition, and, in terms of cobordism, there is no obstruction to the spacetime ending there. No modification of the structure is necessary, which we know is true, since in 5d the $S^1$ is shrinking smoothly into nothing. 

\subsection{Massive type IIA theory}

The massive type IIA (mIIA) \cite{Romans:1985tz} action in the Einstein frame is (for vanishing $F_2$, $F_4$, $H_3$)
\begin{equation}
\label{eq:mIIAaction}
    S\sim \int d^{10}x\sqrt{-g}\big(R -\frac{1}{2}(\partial\phi)^2-\frac{1}{2}e^{\frac{5}{2}\phi} F_0^2\big)  \, ,
\end{equation}
where $F_0=m$ is the Romans mass parameter and $\phi$ the dilaton. While the uncanceled dilaton potential obstructs maximally symmetric 10d solutions, one can write down a solution running in the $x^9$-direction:
\begin{equation}
\label{mIIAsolution}
ds^2=Z(x^9)^{1/12}\eta_{\mu\nu}dx^\mu dx^\nu \, ,\quad e^\phi=Z(x^9)^{-5/6}, \quad Z(x^9)=-3F_0x^9/2\, ,
\end{equation}
where a singularity is encountered at $x^9=0$ and we adopt the conventions of \cite{Calderon-Infante:2026ymy}. It was already noticed in \cite{Buratti:2021yia, Buratti:2021fiv} that this admits a dynamical cobordism description, and in \cite{Angius:2022aeq} the critical exponent of the solution was found to be $\delta=5/\sqrt{2}$. The interpretation of this singularity in the aforementioned papers was as an O8-plane.
In \cite{Calderon-Infante:2026ymy}, this singularity was confronted against both the original Gubser's criterion and against their Geometrization of the Gubser bound $\delta\leq 2\sqrt{\frac{d-1}{d-2}}$. Since $d=10$, both conditions were violated, and the singularity was interpreted as bad. The authors argue that the location of the singularity cannot be where a D8-brane sits, and in fact, a globally well-defined solution would necessitate the introduction of a combination of O8s and D8s, in the spirit of \cite{Apruzzi:2013yva}, hence they conclude that their criterion is right when classifying this singularity as bad.

Our proposal is compatible with the above interpretation, while, at the same time, proposing a concrete guideline for the resolution of the problem. For the action \eqref{eq:mIIAaction}, the bound of \cite{Calderon-Infante:2026ymy} would precisely match our allowed region $\mathcal{R}$, hence we would also get $\delta \notin \mathcal{R}$. The interpretation would then be that the only way the solution can be made compatible with the Cobordism Conjecture is to modify the structure accordingly, i.e., by introducing an appropriate defect. One could make an educated guess and imagine that the defect would need to be the O8-plane, but, in fact, this computation has already been performed in \cite{Blumenhagen:2023abk}, where for the case that the parameters of the generalized Dudas-Mourad action are tuned to be exactly those of the action \eqref{eq:mIIAaction}, they can explicitly find a \emph{charged} End-of-the-World defect with the properties of the O8-plane, i.e., the correct tension and charge. It is crucial that the solution is now charged, i.e., a new higher-form field is included and the structure, and therefore the allowed region, has been modified, so that the singularity can now be considered good, at least with regard to bordism.

This can be done via including a top-form in the action, which consequently modifies the allowed region for $\delta$. Using the results of our appendix \ref{app:TopForm} and taking into account that  $d=10$ and $a=-b_{DM}=-5/2$, in this particular case, the allowed region for $\delta$ is saturated to precisely the single value $\mathcal{R}'={5/\sqrt{2}}$. This is precisely the $\delta$ characterizing both the initial dynamical cobordism solution, as well as the local description of the corresponding defect provided by \cite{Blumenhagen:2023abk}, identified with an O8-plane.

The situation is in perfect agreement with our proposed Sharpened Dynamical Cobordism Conjecture: the initial value of $\delta$ being outside of the allowed range $[0, 3/   \sqrt{2}]$ signals an inconsistency in the initial solution, which we interpret as an obstruction for the initial theory to ``decay'' into nothing, i.e., a non-trivial bordism charge. We could then say that the solution \eqref{mIIAsolution} is in the Swampland. This is not enough to say that massive type IIA (with vanishing $H_3, F_2, F_4$) is automatically in the Swampland, but it signals the presence of a non-trivial bordism group. Within this group, a trivial class exists, which means that one could, in principle, find the ``gauged'' solution where the corresponding bordism charge vanishes, i.e., at least, at the level of bordism there is no fundamental UV-inconsistency present. Once we modify the action to include the top-form, the range of allowed values is extended, and the value of $\delta$ found becomes \emph{good}. This is in direct correspondence with the existence of a corresponding charged defect under the top-form, which now modifies the structure. Schematically, now the solution including the defect carries a trivial bordism charge, and one can interpret the corresponding defect, the O8-plane, as an End-of-the-World defect.

\subsection{The Janis-Newman-Winicour (JNW) solution}
\label{sec:JNW}

Let us now test the Sharpened Dynamical Cobordism against some black hole solutions. Black hole systems have already been considered in a dynamical cobordism framework, focusing mostly on small black holes \cite{Angius:2022aeq, Angius:2023xtu}.

Here, we start with the simplest possible setup one could imagine, simply coupling gravity with a scalar field 
\begin{equation}\label{eq:1USE_actionScalarBH}
    S = \frac{1}{16\pi G} \int d^4x \sqrt{g} \left( R - \frac{1}{2}(\partial \phi)^2\right).
\end{equation}
Then, the Janis-Newman-Winicour (JNW) solution \cite{PhysRevLett20878, Wyman:1981bd} is a static spherically symmetric black hole solution to this theory: 
\begin{subequations}\label{eq:JNW_Solution}
    \begin{equation}
    ds^2 = -f(r)^{\gamma} dt^2 + f(r)^{-\gamma} dr^2 + f(r)^{1 - \gamma} r^2 d\Omega^2_{2}, \qquad  f(r) = 1- \frac{r_-}{r},
    \end{equation}
    \begin{equation}
    \phi(r) = \frac{2\Sigma}{r_-} \ln\left(f(r) \right).
    \end{equation}  
\end{subequations}
Here, \( r_- \) and \( \gamma \) are constants:
\begin{equation}
r_- = 2\sqrt{G^2M^2+\Sigma^2} = \frac{2GM}{\gamma}, \qquad \gamma = \frac{M}{\sqrt{M^2 + \Sigma^2/{G^2} }}.  
\end{equation}
$M$ is the mass and $\Sigma$ is a scalar ``charge'', defined by expanding $\phi$ in the asymptotic limit
\begin{equation}
    \phi = - \frac{2 \Sigma}{r} + O(r^{-2}),
\end{equation}
in analogy to the mass. Both are assumed to be nonzero.  Note that $0 < \gamma < 1$, which implies that $r=r_-$ is a naked singularity since the Ricci scalar is
\begin{equation}
    R = \frac{(1-\gamma^2)r_-^2}{2r^4(1-r_-/r)^{2-\gamma}}.
\end{equation}

 Precisely because of this naked singularity, the Janis-Newman-Winicour solution is widely considered to be unphysical, in accordance with the weak cosmic censorship hypothesis \cite{Penrose:1969, wald1997}. More recently it has been related to negative mass Schwarzschild black holes \cite{Gao:2024lrb}, which are also considered unphysical as, for instance, they can lead to a vacuum whose energy is not bounded from below \cite{Horowitz:1995ta}.
 
 However, we have seen that singularities can be resolved or hidden behind the horizon when considering a more complete description.  
Considering the criteria of section \ref{sec:Bounds}, keeping in mind the limitations in their applicability,  both \eqref{eq:GubserCorolary} and the Gubser potential criterion are satisfied due to the absence of the potential. Applying the computability criterion in this case is unclear to us; as, in addition to lacking a holographic interpretation, \eqref{eq:JNW_Solution} does not coincide with the geometries considered in \cite{Kiritsis:2016kog}. 

However,  we observe that $r_-$ is a point located at finite spacetime distance but infinite field distance, as $\phi$ diverges, so this is a setup compatible with dynamical cobordism. In the limit $r \rightarrow r_-$, for a reference point $c$ with $c>r>r_-$, the spacetime distance to the singularity is
\begin{equation}
    \Delta = \int_r^c f(\lambda)^{\gamma/2} \; d\lambda = \frac{1}{(r-r_-)^{\gamma/2}} \left( \frac{r_-^{\gamma/2}(r-r_-)}{1-\gamma}  \right) +C \sim (r-r_-)^{1-\gamma/2},
\end{equation}
where $C$ depends on $c$. The field distance is 
\begin{equation}
    D = \frac{1}{\sqrt{2}} \left\vert  \phi(r) - \phi(c)  \right\vert = \frac{-1}{\sqrt{2}}\frac{2\Sigma}{r_-} \ln(r-r_-) + C.
\end{equation}
Therefore, $r-r_- \sim e^{- \frac{r_-}{ \sqrt{2}\, \Sigma} D}$, so we can conclude the scaling relations are satisfied with critical exponent
\begin{equation}
    \delta = \frac{GM}{\Sigma} \frac{2}{\sqrt{2}} \left( \frac{2}{\gamma}-1\right) = \frac{2/\gamma -1}{\sqrt{1-\gamma^2}} \frac{2}{\sqrt{2}}.
\end{equation}
Note that $\delta$ depends on the ratio $M/\Sigma$ through $\gamma$. The allowed region for the critical exponent is $ \mathcal{R} =\left[0, 2 \sqrt{\frac{3}{2}}\right]$
hence $\delta \notin \mathcal{R}$ since $\gamma<1$. We interpret this as an indication that the structure of the theory needs to be modified to be rendered compatible with quantum gravity, i.e., the solution as it is should be placed in the Swampland. Note that in this simple case our allowed region is identical to the Geometrization bound of \cite{Calderon-Infante:2026ymy}, however \eqref{eq:JNW_Solution} is not codimension-1, thus the application of such criterion is unclear in this case. 

We consider this a successful test of the Sharpened Dynamical Cobordism, since the value of the critical exponent can be demonstrated to place unphysical setups in the Swampland.

\subsection{The Garfinkle-Horowitz-Strominger (GHS) solution}
\label{sec:GHS}

In contrast to the JNW solution, the Garfinkle-Horowitz-Strominger (GHS) solution \cite{Garfinkle:1990qj, GIBBONS1988741} is a black hole with a singularity hidden behind a horizon. The presence of a Maxwell field carrying charge allows the formation of a horizon in this example. This extra field acts as a new ingredient that may modify the bounds on $\delta$.

We consider the following theory, in Planck units $M_P=1$, 
\begin{equation}\label{eq:1USE_BlackHoles_GHSSolution}
    S = \frac{1}{2}\int d^4x \sqrt{g} \left( R - \frac{1}{2}(\partial \phi)^2 -\frac{1}{4}e^{-\phi}F^2\right),
\end{equation}
which can be regarded as part of the four-dimensional low energy limit of heterotic string theory in the Einstein frame, coming from a compactification from 10 to 4 dimensions and truncating the gauge group, either $E_8 \times E_8$ or $SO(32)$, to a $U(1)$. Here, $F=dA$ is a Maxwell field strength and the scalar field $\phi$ is the four dimensional dilaton. The GHS solution is

\begin{subequations}\label{eq:BlackHole_GHSSolution}
    \begin{equation}
        ds^2 = -f(r) dt^2 + f(r)^{-1} dr^2 + r\left(r-r_-\right) d\Omega^2_2, \qquad f(r) = 1-\frac{r_+}{r},
    \end{equation}
    \begin{equation}
        e^{\phi} = e^{\phi_0} \left(1-\frac{r_-}{r} \right), \qquad\qquad F  = \frac{-q}{r^2}\frac{q^2}{2r_+r_-}dt \wedge dr, 
    \end{equation}
    \begin{equation}
         e^{\phi_0} = \frac{q^2}{2r_+r_-}, \qquad \qquad r_+ = 2M.
    \end{equation}
\end{subequations}
It is specified by the asymptotic value of the dilaton, $\phi_0$, the mass $M$ and the charge $q$, 
\begin{equation}
    q = \frac{1}{4\pi}\int_{S_{\infty}^2} e^{-\phi}*F.
\end{equation}
The Ricci scalar of this geometry is
\begin{equation}\label{eq:CurvatureGHS_1use}
    R = \frac{r_-^2 (r-r_+)}{2 r^3 (r-r_-)^2}.
\end{equation}
As in the previous case, a singularity is located at $r_{-}$ and physically there is no extension beyond that point. As for $r_+$, it is just a coordinate singularity as long as $r_+>r_-$. This is the location of the horizon, and the solution is continued beyond $r_+$ until reaching the singularity located at $r_-$.

As in the JNW solution, $r_-$ is located at a finite spacetime distance and infinite field distance. Assuming that $r_+ \gg r_-$, when approaching $r_-$ the spacetime distance becomes
\begin{equation}\label{eq:eq:STDistanceGHS_1use}
    \Delta = \sqrt{c \left|c-r_+\right|}-\sqrt{r \left|r-r_+\right|}+r_+ \Big( \tanh ^{-1}\left( |f(c)|\right)- \tanh ^{-1}\left(|f(r)|\right) \Big) \sim r-r_-  .
\end{equation}
The moduli distance is
\begin{equation}\label{eq:ModuliDisGHS_1use}
    D = \frac{-1}{\sqrt{2}}  \ln\left(\frac{1-r_-/r}{1-r_-/c}\right), \quad \longrightarrow \quad  r-r_- \sim e^{-\sqrt{2}D}. 
\end{equation}
Therefore, $\delta=2\sqrt{2}$, independent of the parameters of the solution. 

Since the action is directly of the form examined in section \ref{sec:HigherForm}, taking into account that now $\phi \to - \infty $ and $a<0$ (so effectively it corresponds to the case $\phi \rightarrow+\infty$ and $a>0$ of section \ref{sec:HigherForm}) we find that the allowed region is $\mathcal{R} =\left[0, 2 \sqrt{\frac{3}{2}} \right]$, and is in fact unchanged, despite the presence of the higher form. 
Consequently, $\delta \notin \mathcal{R}$. This seems puzzling, since the solution can be embedded in heterotic string theory \cite{Garfinkle:1990qj}. However, the singularity is hidden behind a horizon, it is not a naked singularity. As Gubser criteria for naked singularities were used to find $\mathcal{R}$, together with the fact that $dr$ is a timelike direction inside the horizon, we are agnostic about the applicability of Sharpened Dynamical Cobordism. We will therefore look at the extremal solution, which is the case where the value of $\delta$ can lead to a trustworthy analysis.

This corresponds to $r_-=r_+$. Now there is a naked singularity at $r_+$, since $ R \sim \frac{1}{r-r_+}$ and the spacetime distance scales like
\begin{equation}
    \Delta \sim 2 \sqrt{r_+}\; \sqrt{r-r_+}
\end{equation}
The moduli distance is the same as in \eqref{eq:ModuliDisGHS_1use}, therefore, the value of $\delta$ has changed to be $ \delta = \sqrt{2} \in \mathcal{R}$. Note that this value is precisely the one given by $\delta_{00}$, the orange curve in the right side of figure \ref{fig:DeltaAndExponentConstraints}. 

All in all, in the extremal version, we have a naked singularity that, locally, satisfies the Sharpened Dynamical Cobordism bounds. In this case the bound happens to coincide with that of an ED theory, i.e. is numerically also the same as the Gubser geometrization bound of \cite{Calderon-Infante:2026ymy} . Similarly, both versions of the original Gubser criterion are also automatically satisfied because the geometry is the extremal limit of a family of black hole solutions.

Now, returning to our bound, the fact that $\delta\in \mathcal{R}$ indicates that the singularity could be resolved within the theory we are considering. This agrees with the literature; for instance, in \cite{Garfinkle:1990qj}, it was shown how the extremal magnetic dual version of \eqref{eq:BlackHole_GHSSolution} is regular in the string frame, showing how the solution has no problems when considered in string theory. In addition, \eqref{eq:BlackHole_GHSSolution} can be uplifted to a five dimensional background to all orders in $\alpha'$, representing an exact string solution \cite{Horowitz1994ExtremalBH}.

Interestingly, we can also study the naked singularity that arises in the super-extremal limit, $r_- >r_+$. The computations are the same as in the sub-extremal one, then
\begin{equation}
    \delta= 2\sqrt{2} \notin \mathcal{R}.
\end{equation}

In agreement with our expectations for super-extremal black holes, this value of $\delta$ indeed corresponds to the disallowed region of the Sharpened Dynamical Cobordism.

To end this section, we would like to remark how the addition of a gauge field has influenced the singularity. In the JNW solution, we obtained a $\delta$ exponent which was not in $\mathcal{R}$. We can think of the GHS solution as adding a gauge field. With this extra structure, which modifies $\Omega$ because of the presence of the gauge bundle, $\mathcal{R}$ changes. At the end, the extremal version has a $\delta$ in the region, indicating that the resolution of the singularity should be possible with the field content of the solution.

\subsection{Continuous distribution of D3-branes}
\label{sec:D3continuous}

A continuous distribution of D3-branes could give rise to
problematic singularities. Already in \cite{Gubser:2000nd} it was observed that the $\sigma_5$ distribution of \cite{Freedman:1999gk} violates the Gubser potential criterion. 

Let us briefly describe the system at hand. The geometry is a solution to type IIB supergravity, and it represents a continuum distribution of D3-branes given by $\sigma$
\begin{equation}
    ds^2 = H^{-1/2} (-dt^2 + d\mathbf{x}^2) + H^{1/2} \delta_{ab} dy^a dy^b, \qquad\qquad H = \int d^6\xi \; \sigma(\xi) \frac{L^4}{|\mathbf{y} - \mathbf{\xi}|^4}\, ,
\end{equation}
where the possible distributions can be found in \cite{Freedman:1999gk}. It can be truncated on an $S^5$ to give a solution to $d=5$ $\mathcal{N}=8$ supergravity involving the metric and scalars in the $SL(6, \mathbb{R})/SO(6)$ coset, which, when it exists, is dual to the Coulomb branch of $\mathcal{N}=4$ super Yang-Mills. 

The solution solves the EOMs of the following action 
\begin{equation}
    S \sim \int d^5x \sqrt{g} \left( R - \frac{1}{2}\sum_{i=1}^5 (\partial\alpha_i)^2 - V\right),
\end{equation}
where it is convenient to define new fields $\beta_a$ as
\begin{equation}
\begin{pmatrix}
\beta_1 \\
\beta_2 \\
\beta_3 \\
\beta_4 \\
\beta_5 \\
\beta_6
\end{pmatrix}
=
\begin{pmatrix}
1/\sqrt{2} & 1/\sqrt{2} & 1/\sqrt{2} & 0 & 1/\sqrt{6} \\
1/\sqrt{2} & -1/\sqrt{2} & -1/\sqrt{2} & 0 & 1/\sqrt{6} \\
-1/\sqrt{2} & -1/\sqrt{2} & 1/\sqrt{2} & 0 & 1/\sqrt{6} \\
-1/\sqrt{2} & 1/\sqrt{2} & -1/\sqrt{2} & 0 & 1/\sqrt{6} \\
0 & 0 & 0 & 1 & -\sqrt{2/3} \\
0 & 0 & 0 & -1 & -\sqrt{2/3}
\end{pmatrix}
\begin{pmatrix}
\alpha_1 \\
\alpha_2 \\
\alpha_3 \\
\alpha_4 \\
\alpha_5
\end{pmatrix},
\end{equation}
constrained by $\sum \beta_a=0$. Thanks to this definition the potential, whose exact form can be checked at \cite{Freedman:1999gk}, is proportional to
\begin{equation}
    V \propto (\Tr(M))^2-\Tr(M^2),
\end{equation}
where $M= \mathrm{diag}(\{e^{2\beta_a}\}_{a=1}^6)$. The geometry is
\begin{equation}\label{eq:DistributionOfBranes_5DSolution}
    ds^2=e^{-2\sigma(\rho)} (-dt^2 + d\mathbf{x}^2) + d\rho^2, \qquad \alpha_i = \alpha_i(\rho).
\end{equation}
There are five possible options that represent different distributions of D3-branes, each characterized by a six-dimensional vector $\gamma_a$. 

We are interested in the region close to the singularity ($\mu \rightarrow \infty$), where we have 
\begin{equation}
    \beta_a = 2\gamma_a \mu(\rho), \qquad \frac{d\mu}{d\rho} = \frac{-g}{4} \sum_{a=1}^6\gamma_a e^{2\gamma_a \mu}, \qquad \frac{d \sigma}{d\rho} = \frac{-g}{12}  \sum_{a=1}^6 e^{2\gamma_a \mu},
\end{equation}
where $g$ is the coupling of the theory.

The problematic distribution is given by \cite{Gubser:2000nd, Freedman:1999gk}
\begin{equation}\label{eq:D3BranesDistribution_Problematic}
\sigma_5(\xi) = \frac{1}{\pi^{3/2} \ell^2} \left( -\frac{1}{2} \frac{1}{(\ell^2 - \xi^2)^{3/2}} \, \theta(\ell^2 - \xi^2) + \frac{1}{\sqrt{\ell^2 - \xi^2}} \, \delta(\ell^2 - \xi^2) \right),
\end{equation}
where $\ell$ is a length scale that represents the size of the distribution. According to \cite{Gubser:2000nd, Freedman:1999gk}, this distribution involves ghost D3-branes with negative tension whenever $\sigma_5 < 0$, without an explanation for these objects, it should be regarded as unphysical. 

Close to the singularity we have
\begin{equation}
    \frac{d\mu}{d\rho} = -\frac{g \sqrt{5}}{4\sqrt{3}} e^{\frac{10}{\sqrt{15}} \mu}, \qquad \sigma = \frac{\mu}{\sqrt{15}}, \qquad (\gamma_a) = \frac{1}{\sqrt{15}}(5,-1,-1,-1,-1,-1).
\end{equation}
The spacetime distance is immediately $\Delta \sim \rho$, while for the field distance in general we have
\begin{equation}
\begin{aligned}
    dD^2 &= 2\left[ \gamma _2^2+\gamma _3^2+\gamma _4^2+\gamma _5^2+\gamma _6^2+ \right.\\
     &+ \left.\gamma _4 \gamma _5+\left(\gamma _3+\gamma _4+\gamma _5+\gamma _6\right) \gamma _2+\left(\gamma _4+\gamma _5\right) \gamma _6+\gamma _3 \left(\gamma _4+\gamma _5+\gamma _6\right)\right] d\mu^2.
\end{aligned}
\end{equation}
In our case,  $D \sim \sqrt{2}\; \mu$. The curvature is
    $R = 4 \left( 2 \sigma''(\rho) - 5 \sigma'(\rho)^2\right)$,  so we
 find that the scaling relations are satisfied with  $\delta = 2\sqrt{\frac{10}{3}}.$
The allowed region is in this case given by $\mathcal{R}=[0,\delta_0]=[0,\frac{4}{\sqrt{3}}]$, therefore $\delta \notin \mathcal{R}$, as expected. This is in agreement with the Gubser potential and horizon criteria, because $V \rightarrow +\infty$ near the singularity at $\mu\rightarrow +\infty$ \cite{Gubser:2000nd}. 

The same process can be used  for the four other distributions of branes in \cite{Freedman:1999gk}. The results are summarized in table \ref{tab:DeltasOfBraneDistributions} and we note that in all remaining cases  $\delta \in \mathcal{R}$. This is in agreement with the Gubser potential criterion as $V \rightarrow -\infty$ at the singularity and with the Geometrization of \cite{Calderon-Infante:2026ymy}. Once again, the Sharpened Dynamical Cobordism gives a result consistent with expectations and can be used to differentiate between good and bad singularities.

\begin{table}[h]
\centering
\begin{tabular}{c| c| c}
\toprule
$\sigma_i(\xi)$ & $\gamma_a$ & $\delta/2$ \\
\hline
$\sigma_1$ & $(1,1,1,1,1,-5)/\sqrt{15}$ & $\sqrt{2/15}$ \\ \midrule
$\sigma_2$ & $(1,1,1,1,-2,-2)/\sqrt{6}$ & $\sqrt{1/3}$ \\ \midrule
$\sigma_3$ & $(1,1,1,-1,-1,-1)/\sqrt{3}$ & $\sqrt{2/3}$ \\ \midrule
$\sigma_4$ & $(2,2,-1,-1,-1,-1)/\sqrt{6}$ & $2\sqrt{1/3}$ \\ \midrule
$\sigma_5$ & $(5,-1,-1,-1,-1,-1)/\sqrt{15}$ & $\sqrt{10/3}$ \\
\bottomrule
\end{tabular}
\caption{Critical exponents associated with each of the 5d D3-brane distributions $\sigma_i$ of \cite{Freedman:1999gk}.}
\label{tab:DeltasOfBraneDistributions}
\end{table}

These solutions only contain scalar fields and have known holographic duals, the Coulomb branch of $\mathcal{N}=4$ SYM. Therefore, the computability bound of \cite{Kiritsis:2016kog} is applicable, which translated in the dynamical cobordism language implies
\begin{equation}
    \delta^{Co} < 2 \sqrt{\frac{2}{3}},
\end{equation}
as we have seen in section \ref{sec:boundsDC}.
The admittedly problematic distribution $\sigma_5$ violates this bound. We observe that the first two distributions $\sigma_1$ and $\sigma_2$ satisfy it, while $\sigma_3$ marginally violates it, since it corresponds to $\delta=\delta^{Co}$. Similarly, $\sigma_4$ at the level of Sharpened Dynamical Cobordism presents no issues, but it clearly violates the computability bound. It would be interesting to study these cases in more detail and observe what exactly fails when computing the perturbation spectrum and try to observe whether the holographic duals are well-defined.

\subsection{Repulson singularities}
\label{sec:Repulson}

An interesting class of singularities are the so-called repulson singularities \cite{Behrndt:1995tr,Kallosh:1995yz,Cvetic:1995mx}, which can be realized in certain black hole solutions of heterotic string theory. The name comes from the fact that massive particles experience an infinite repulsive potential at the singularity \cite{Kallosh:1995yz}. This means that if the singularity is located at $r=0$, there is a minimum radius $r_{min}$ that a particle can reach before bouncing. In general $r_{min}$ may depend on the mass or angular momentum of the incoming particles. 
Some repulson singularities can be resolved by the mechanism of \cite{Johnson:1999qt}, and thus they are a suitable setup to test Sharpened Dynamical Cobordism.

\subsubsection{A D2-D6 brane system and the enhançon geometry}

Let us first describe the D2-D6 system in which the repulson singularity arises. Both branes extend along the $045$ directions while the $D6$ is also wrapping a $K3$.

The solution in the Einstein frame is \cite{Johnson:1999qt}
\begin{equation}\label{eq:D2D6System}
\begin{aligned}
ds^2 &= Z_2^{-5/8} Z_6^{-1/8} \eta_{\mu \nu} dx^\mu dx^\nu + Z_2^{3/8} Z_6^{7/8} ds^2_3 + V^{1/2} Z_2^{3/8} Z_6^{-1/8} ds^2_{K3}, \\
e^{2\phi} &= Z_2^{1/2} Z_6^{-3/2}, \qquad F_4 = d(Z_2^{-1}) dx^0 \wedge dx^4 \wedge dx^5, \qquad Z_2 = 1-\frac{r_2}{r}, \\
F_8 &= d(Z_6^{-1}) dx^0 \wedge dx^4 \wedge dx^5 
\wedge dx^6 \wedge dx^7 \wedge dx^8 \wedge dx^9, \qquad Z_6 = 1+\frac{r_6}{r}.
\end{aligned}
\end{equation}
$V$ is the volume of a $K3$ and $ds_{K3}$ is the metric on a unit-volume $K3$. $\mu$, $\nu$ describe the 045 directions shared by all branes and $ds^2_3$ has the directions transverse to all of them
\begin{equation}
    ds^2_3 = dr^2 + r^2 \left( d\theta^2 + \sin^2(\theta)d\varphi^2\right).
\end{equation}
The $K3$ is tangent to $D6$-branes, but transverse to the $D2$-branes. The radii $r_2$, $r_6$ for a stack of $N$ branes are \cite{Johnson:1999qt}
\begin{equation}
    r_2 = N\frac{(2\pi)^4 g_s \alpha'^{5/2}}{2V}, \qquad r_6 = N\frac{g_s \alpha'^{1/2}}{2}.
\end{equation}
\eqref{eq:D2D6System} is a solution to the EOMs coming from the action
\begin{equation}
    S = \frac{1}{16\pi G_{10}} \int d^{10}x \sqrt{g} \left( R - \frac{1}{2}(\partial \phi)^2 - \frac{1}{2\cdot 4!} e^{\phi/2}F_4^2 - \frac{1}{2\cdot 8!} e^{-3\phi/2}F_8^2\right).
\end{equation}
Note that in the usual $D2$-$D6$ system $Z_2=1+r_2/r$, unlike in \eqref{eq:D2D6System}. Now, $Z_2$ vanishes at $r=r_2$, which is a repulson (naked) singularity, as can be observed in the Ricci scalar
\begin{equation}
\begin{aligned}
        R&=\frac{\left(21 r_6^2-3 r_2^2+14 r_6 r_2\right) r^2-4 r_2 r_6 \left(5 r_2+7 r_6\right) r+4 r_2^2 r_6^2}{64 r^2 \left(r-r_2\right){}^2 \left(r+r_6\right){}^2} \\[1.5mm]
        &= \frac{-3}{64 (r-r_2)^2} + O\left( \frac{1}{r-r_2} \right).
\end{aligned}
\end{equation}

Despite the singular behavior, the geometry \eqref{eq:D2D6System} is argued to be physically realized, for example, as the S-dual of a combination of the Kaluza-Klein and $H$ monopoles in heterotic string theory \cite{Khuri:1992ww, Johnson:1999qt, Johnson:1995bf}.

The singularity can be probed by $D6$-branes wrapped on the $K3$. One would think that, as the system is supersymmetric the probe is free to move until it reaches $r_2$.  However, \cite{Johnson:1999qt} pointed out an obstruction to this process: The action of a D6 probe brane in the presence of the D2-D6 system expanded in the transverse velocity $\dot{\mathbf{\xi}}^2$ is
\begin{equation}\label{eq:1USE_MetricProbe}
    S_{probe} \sim \int d^3\xi \; \frac{1}{2}\left( (4\pi)^{-4}\alpha'^{-2}VZ_2-Z_6\right) \dot{\mathbf{\xi}}^2  + ...
\end{equation}

The coefficient for $\dot{\xi}^2$ changes sign at some finite $r_e> r_2$.
The probe becomes tensionless at $r_e$ and its tension becomes negative in the region $r_2<r<r_e$. To solve this problem, the absolute value must be taken at the original DBI action, hence the probe will feel a potential after crossing $r_e$, thus probe D6-branes wrapped on the K3 are stuck at this radius \cite{Johnson:1999qt}.

This argument demonstrates an obstruction to reaching $r_2$. In fact, the probe becomes tensionless at $r_e$ and at the same time it expands and melts into a shell of $N$ monopoles \cite{Johnson:1999qt}.
This is because, at the same time a wrapped D6 becomes tensionless, a wrapped D4 becomes massless. 

The D6-branes of the solution act as a source for a RR 4-form in 6d,
but this form is dual to a 2-form RR field under which the D4 branes are charged. Together with an anti-D4 and a RR vector, an enhanced $SU(2)$ symmetry is formed. For $m$ the mass of the boson of the wrapped D4s, the sourced 2-forms create a monopole of spatial extension $m^{-1}$. Close to $r_e$ $m \rightarrow0$, so the monopole extension increases until filling the whole shell located at $r_e$ \cite{Johnson:1999qt}.

Now, the geometry \eqref{eq:D2D6System} is only valid until $r_e$. Beyond $r_e$ the geometry is flat and it is obtained by substituting $r=r_e$ in \eqref{eq:D2D6System}~\cite{Johnson:1999qt}. Thus, the singularity disappears. Note that no additional field was necessary to account for the resolution.
 In the interior region, $r<r_e$, there is an unbroken $SU(2)$ gauge symmetry, and thus this is called enhançon geometry \cite{Johnson:1999qt}.

\subsubsection{Probing the singularity}

Let us try to confront the singularity at $r_2$ with the Sharpened Dynamical Cobordism. First we need to verify the scaling relations hold. Indeed, $r_2$ is located at finite spacetime distance but infinite field space distance,

\begin{equation}
    \Delta = \int_r^c dx \left( 1- \frac{r_2}{x} \right)^{3/16} \left( 1+ \frac{r_6}{x} \right)^{7/16} \sim (r-r_2)^{19/16},
\end{equation}
and using the field distance $D = \frac{1}{\sqrt{2}} \vert \phi(c) - \phi(r)\vert$, we obtain $   r-r_2 \sim e^{4\phi} $.
Thus, the scaling relations are recovered with the coefficient
\begin{equation}
    \delta = \frac{19 \sqrt{2}}{2}.
\end{equation}
To find  $\mathcal{R}$, note that the singularity comes from $Z_2$ going to 0, so $F_4$ is divergent, so we can use our results in \eqref{eq:DeltaBoundsHigherForm}. As $\phi \rightarrow- \infty$, $\mathcal{R}_a$ is given by substituting $a=-1/2$, $q=4$ and $d=10$:
\begin{equation}
    \mathcal{R}= \left[ 0, \frac{3 \sqrt{2}}{2} \right] \cup \left[\frac{19\sqrt{2}}{2}, +\infty \right]
\end{equation}
Therefore, the $r_2$ singularity is in $\mathcal{R}$. 
The solution lacks a potential, so it satisfies the Gubser potential criterion and \eqref{eq:GubserCorolary} automatically.  
However, we observe $\delta > \delta_0$, i.e., a value outside the region corresponding to an ED theory. This illustrates the importance of taking the additional fields into account for the determination of the allowed region. If we had ignored their presence, then the Sharpened Dynamical Cobordism interpretation would have been that the singularity is problematic, which is not the case, since an explicit resolution within the same theory is provided.

\section{Concluding Thoughts}
\label{sec:conclusions}

In the present paper we have proposed a sharpened version of dynamical cobordism which places certain bounds on the values of the critical exponent $\delta$ in order to ensure compatibility with a theory of quantum gravity, interpreting the solutions with ``bad'' values of $\delta$ as solutions where an obstruction for the theory to decay to nothing exists, i.e., the solution corresponds to a non-trivial cobordism charge. Unlike the Sharpened Distance Conjecture, where the bound for the mass decay coefficient only depends on the number of spacetime dimensions, for determining the allowed region for the critical exponent, one needs to take into account more physical input, such as the fields included in the action and their couplings. Hence, we interpret this allowed region $\mathcal{R}^\xi$ as structure-dependent.

Consequently, we propose that these singular solutions with $\delta \notin \mathcal{R}^\xi$ should be placed in the Swampland, unless a certain modification of the cobordism structure is performed, such that now $\delta \in \mathcal{R}^{\xi ' }$. A possible realization of this is the inclusion of new defects; hence, dynamical cobordism can now be used to pre- and postdict stringy defects. The value of the critical exponent and the corresponding scaling relations can then be used as a guideline for providing an explicit spacetime description for the new defect, in the spirit of \cite{Blumenhagen:2022mqw, Blumenhagen:2023abk}, but this is not an avenue explored in the present paper. 

For this version of the conjecture to be of any use, one needs a systematic way to determine the allowed region. Here, we propose a Gubser-inspired criterion, keeping in mind that it should now be generally applicable, and not only in holographic setups. We note that in the simple case of a codimension-1 singular solution of a scalar theory exhibiting the dynamical cobordism scalings, the proposed allowed region is only dimension-dependent, with $\delta \leq 2 \sqrt{\frac{d-1}{d-2} }$. This is compatible with the recently proposed geometrization of Gubser proposed in \cite{Calderon-Infante:2026ymy}. However, once we allow higher-codimension defects and/or the inclusion of additional fields in the action, i.e., by modifying the structure, additional allowed regions open up. This means that, taking the new structure into account, an initially ``bad'' solution can now become ``good''. We note that for the Gubser-inspired determination of $R_a$, the converse does not happen, i.e., the inclusion of extra structure never turns an initially ``good'' solution into a ``bad'' one, at least in the example we explicitly studied. It remains to be seen whether this is a general feature or a mere coincidence.

We have tested our version of the Conjecture against several examples, some of which were newly identified as dynamical cobordisms, summarized in table \ref{tab:summary}. For completeness, we have also shown the compliance or not of the same solutions with the other criteria for singularities, which have been reviewed in section \ref{sec:AllLiterature}.  We find it remarkable how our proposed bound is in complete agreement with expectations from string theory, i.e., it places in the Swampland solutions we truly expect to be bad, while it is non-trivially compatible with solutions that are good. We find that the case of massive IIA \emph{with} the O8-plane is a particularly nice illustration of the situation, since the solution would have been disallowed without the structure-dependent modification of the allowed region, as actually happened according to \cite{Calderon-Infante:2026ymy}.

\begin{table}[!htbp]
\centering
\resizebox{\textwidth}{!}{%
\begin{tabular}{l|c|c|c|c|c}
\toprule
   &  \multicolumn{5}{c}{Criteria}  \\
     Solution  & Gubser potential\cite{Gubser:2000nd}        & Gubser Horizon\cite{Gubser:2000nd}       & Computability\cite{Kiritsis:2016kog}     & Geometrization\cite{Calderon-Infante:2026ymy}       &  Sharpened DC       \\
\midrule
BoN \cite{Witten:1981gj} & \checkmark & ? & - & \checkmark & \checkmark \\
mIIA (w/o O8) \cite{Angius:2022aeq} & \ding{55} & \ding{55} & - &  \ding{55} & \ding{55} \\
mIIA w/ O8 \cite{Blumenhagen:2023abk} & \ding{55} & \ding{55} & - & \ding{55} & \checkmark \\
JNW BH \cite{PhysRevLett20878}  & \checkmark & \checkmark & - & - & \ding{55} \\
GHS BH \cite{Garfinkle:1990qj} & \checkmark & \checkmark & - & - & \checkmark \\
$\sigma_1, \sigma_2$ D3-distribution \cite{Freedman:1999gk} & \checkmark & \checkmark & \checkmark & \checkmark & \checkmark \\
$\sigma_3, \sigma_4$ D3-distribution \cite{Freedman:1999gk} & \checkmark & \checkmark & \ding{55} & \checkmark & \checkmark \\
$\sigma_5$ D3-distribution \cite{Freedman:1999gk} & \ding{55} & \ding{55} & \ding{55} & \ding{55} & \ding{55} \\
D2-D6 repulson \cite{Johnson:1999qt} & \checkmark & ? & - & - & \checkmark \\
\bottomrule
\end{tabular}
}
\caption{Evaluation of solutions across criteria reviewed in this paper and the Sharpened Dynamical Cobordism}
\label{tab:summary}
\end{table}

There are multiple intriguing future directions one could pursue.  To begin with, we would like to clarify to what extent our Gubser-motivated method for determining the allowed region for $\delta$ is applicable, and to work out the explicit regions for more complicated actions. It would be interesting to find other physical arguments supporting the selection of the ``good'' delta regions, possibly without any holographic origin. The finiteness of the action explored in appendix \ref{app:finite_action} provides such a hint.

Once the allowed regions for more complicated actions are determined, we can test more complicated systems. Setups particularly suitable for such an exploration are the flux-backtracked solutions of \cite{Apers:2025pon}, which exhibit dynamical cobordism scalings. If our expectation that the inclusion of suitable extra structure brings us closer to the true quantum gravity structure $\Omega^{QG}=0$, i.e., the allowed regions for $\delta$ become wider, then solutions with  $\delta \leq 2\sqrt{\frac{4-1}{4-2}}=\sqrt{6}$ would respect the Sharpened Dynamical Cobordism Conjecture. The flux-backtracked solution corresponding to scale-separated DGKT \cite{DeWolfe:2005uu} corresponds to $\delta=\frac{3\sqrt{6}}{\sqrt{13}}$\footnote{We thank Fien Apers for pointing this out to us and for related enlightening discussions.}\cite{Bedroya:2025ltj}, hence would most likely be compatible with Sharpened Dynamical Cobordism. Given the relation of these setups and those of \cite{Bedroya:2025ltj} to holography, the use of the Gubser-inspired allowed region is even more strongly justified.

Another direction in which the Sharpened Dynamical Cobordism could yield useful insights into the physical interpretation of singular solutions is within the framework of non-supersymmetric string theories. We remind the reader that the cobordism framework is completely independent of supersymmetry and is therefore particularly suited for such endeavors. For instance, \cite{ Mourad:2024dur} has explored some vacua of $SO(16) \times SO(16)$\cite{Alvarez-Gaume:1986ghj} and $USp(32)$\cite{Sugimoto:1999tx} strings, and we hope the associated values of $\delta$ would offer some hints about the resolution of the singularities. Similarly, it would be interesting to observe the interplay of the sourced solutions of \cite{Raucci:2022jgw} with dynamical cobordism.

In a different direction, one would try to establish what is wrong with the singularities that are not compatible with a transition to nothing. It would be very interesting to examine whether the allowed region can be influenced by other properties of the theory, such as chirality, see for instance the constructions of \cite{Altavista:2026edv, Angius:2024pqk}.

Additionally, we find it very important to understand the interrelations between Sharpened Dynamical Cobordism and other Swampland Conjectures. This is, however, not trivial, since for dynamical cobordism one often deals with a field space rather than the moduli space; hence, the potential would necessitate appropriately modified versions of other conjectures, especially the Distance Conjecture. While such attempts have been made \cite{Basile:2023rvm, Mohseni:2024njl, Debusschere:2024rmi, Demulder:2024glx, Demulder:2026cfo}, there is no clear consensus.
Despite this, there are indications of such connections: For instance, in \cite{Raucci:2026fzp}, the Sharpened Distance Conjecture  in a warped compactification scenario was found to constrain the growth of the potential, which can in turn be related to the critical exponent. Similarly, the Sharpened Distance Conjecture was related to the critical exponent through the AdS Distance Conjecture \cite{Lust:2019zwm} in \cite{Blumenhagen:2022mqw}. We feel a careful exploration of all these relations is worthwhile.

Finally, we would like to test more exotic setups using the Sharpened Dynamical Cobordism. It is interesting to check whether the singularities in \cite{Dijkgraaf:2016lym, Medevielle:2023jmn} respect the conjecture and what the physical interpretation of such a cobordism transition would be.  We will explore these questions in an upcoming paper \cite{Makridou:2026ex}.

\vspace{0.5cm}

\noindent
\subsubsection*{Acknowledgments}

{\sloppy\emergencystretch=2em   
The work of AM and AP is supported through the grants
CEX2020-001007-S, PID2021-123017NB-I00 and PID2024-156043NB-I00,
funded by MCIN/AEI/10.13039/501100011033, and ERDF, EU.
The work by AP is supported by the FPU grant no.\ FPU24/02890
from the Spanish Ministry of Science, Innovation and Universities.\par}

We would like to thank  
F. Apers, B. Valeixo Bento, V. Collazuol, N. Cribiori, B. Fraimann, E. Kiritsis, C. Kneissl, L. Paoloni, A. Paraskevopoulou, 
T. Raml, S. Raucci and A. Uranga for insightful discussions about this project. We are indebted to N. Cribiori and S. Raucci for useful comments on this draft. We also want to thank Jos{\'e} Calder{\'o}n-Infante for thoughtful discussions related to the present project and their recent work \cite{Calderon-Infante:2026ymy}.



\appendix

\section{Dimensional reduction to a codimension-1 ansatz}
\label{app:DimensionalReduction}
 
In this appendix we provide a complementary way of obtaining the allowed region $\mathcal{R}$ in the presence of the higher-form field. In particular, we dimensionally reduce the theory \eqref{eq:Actino_ScalarQForm} such that it becomes compatible with a codimension-1 dynamical cobordism solution, as in \cite{Angius:2022aeq}, and examine what bounds can then be imposed on $\delta$\footnote{We thank José Calderón-Infante for pointing out to us the possibility of performing this compactification and dualization procedure.},

We start in $d$ dimensions and
compactify on a $(d-q)-$dimensional manifold, according to the Ansatz 
\begin{equation}
    ds^2_d = e^{-2\alpha \omega} ds^2_q + e^{2 \beta \omega} ds^2_{d-q}. 
\end{equation}
Here $\omega$ is the radion, and $\alpha$ and $\beta$ are chosen to ensure the low-energy effective action is in the Einstein frame and the  radion is canonically normalized:
\begin{equation}
    \alpha = \frac{d-q}{q-2} \beta, \qquad \beta^2= \frac{1}{2} \frac{q-2}{(d-q)(d-2)}.
\end{equation}

For simplicity, we consider the internal manifold to be flat. Then, the lower-dimensional action is 
\begin{equation}
    S \sim \int d^qx \sqrt{g} \left( R - \frac{1}{2}(\partial \phi)^2 - \frac{1}{2}(\partial \omega)^2 - e^{-\frac{2(d-q)}{q-2} \beta \omega} V(\phi) -\frac{1}{2q!}e^{a\phi+\frac{2(d-q)(q-1)}{d-2}\omega} F_q^2\right),
\end{equation}
where $F_q$ is a top-form, i.e., $F_q = f(x) \sqrt{g}\, d^qx$. Additionally, its equation of motion is algebraic: 
\begin{equation}
    0=d * e^{a\phi+\frac{2(d-q)(q-1)}{d-2}\omega}F_q = d\left(- f(x)\; e^{a\phi+\frac{2(d-q)(q-1)}{d-2}\omega} \right).
\end{equation}
It can be solved exactly  
\begin{equation}
    f(x) = -\,e^{-a\phi-\frac{2(d-q)(q-1)}{d-2}\omega}\; Q,
\end{equation}
 where $Q$ is a constant. Substituting in the equations of motion we identify a contribution to the potential, and the effective action can be brought in the form

\begin{equation}\label{eq:LowerDimensionalAction}
    S \sim \int d^qx \sqrt{g} \left( R - \frac{1}{2}(\partial \phi)^2 - \frac{1}{2}(\partial \omega)^2 - e^{-\frac{2(d-q)}{q-2} \beta \omega} V(\phi) -\frac{Q^2}{2}e^{-a\phi-\frac{2(d-q)(q-1)}{d-2}\omega}\right).
\end{equation}

Now we follow the same procedure as in section \ref{sec:ScalarDynamicalCobordism} to find the bounds for $\delta_V$. As promised, this is now compatible with a codimension-1 solution
\begin{equation}
    ds^2_q = e^{-2\zeta(z)}ds^2_{q-1} +dz^2.
\end{equation}
The ansatz is
\begin{equation}\label{eq:Ansatz2ScalarFields}
    V = C_V e^{\delta_V \phi}, \qquad \phi = -C_{\phi} \ln(z)+\ln(M_{\phi}), \qquad \omega = -C_{\omega} \ln(z), \qquad \zeta = C_{\zeta} \ln(z).
\end{equation}
Note that we parametrize $V$ in the same way as \eqref{eq:HigherFormLogarithmicAnsatz} with the aim of comparing with \eqref{eq:DCGaugeSolution}, but this is not the critical exponent of the lower-dimensional effective action. In fact, the critical exponent is given by
\begin{equation}
    \delta_{EFT} = \frac{2 \sqrt{2}}{\sqrt{C_{\phi}^2 + C_{\omega}^2}}.
\end{equation}

The equations of motion are analogous to \eqref{eq:EOMs_EDTheory}. If we substitute \eqref{eq:Ansatz2ScalarFields} in them, we have again three cases depending on which one potential term dominates. 
Here we present the explicit results for the case we deem most interesting,  where both contributions to the potential are significant, i.e.,  both potential terms grow as much as the curvature and kinetic terms. This implies

\begin{equation}
    C_{\phi} = \frac{2 (q-2)}{a+(q-1) \delta _V}, \qquad  C_{\omega} = -\frac{(q-2) \left(a+\delta _V\right)}{\beta  (d-q) \left(a+(q-1) \delta _V\right)}.
\end{equation}
Solving for $C_{\zeta}$ in the Einstein equations and for $C_V$ and $Q^2$ in the scalar EOMs, we note the remaining equation is trivially satisfied. The final solution is
\begin{equation}
\begin{aligned}
    C_V &= -c\frac{(q-2)^2 \Xi_1 \Xi_2}{(d-q)^2 \left(a+(q-1) \delta _V\right){}^4}, \qquad Q^2 = 2c^{-a/\delta_V}\frac{(q-2)^2 \Xi_0 \Xi_2}{(d-q)^2 \left(a+(q-1) \delta _V\right){}^4}, \\[2mm]
    C_{\zeta} &= -\frac{(a+\delta_V)^2+ 2 (q-2) (d-q)}{(d-q) \left(a+(q-1) \delta _V\right){}^2}, \qquad \qquad c = M_\phi^{-\delta_V} >0.
\end{aligned}
\end{equation}
We observe that this gives results identical to \eqref{eq:DCGaugeSolution} when imposing $C_V \le 0$ and $Q^2\ge0$ as a consistency condition. The only quantities allowed to change sign are the $\Xi_i$ functions, exactly as in \eqref{eq:DCGaugeSolution}.

One can similarly show that the subleading cases completely reproduce the analysis of section \ref{sec:HigherForm}. 

This supports our generalization of the codimension-1 Ansatz \eqref{eq:Generalized_Ansatz} and our interpretation of the scaling relations as a key property of dynamical cobordisms. In addition, note that this compactification procedure is not valid for $q=2$, while the procedure of section \ref{sec:HigherForm} is.

\section{Allowed region for the case of a top form}
\label{app:TopForm}
Here, we compute the allowed region $\mathcal{R}^{d}$ for a theory with a top-form
\begin{equation}
    S \sim \int d^dx \sqrt{g} \left( R - \frac{1}{2}(\partial \phi)^2-V(\phi)-\frac{1}{2d!}e^{a\phi}F_d^2\right).
\end{equation}
As explained in appendix \ref{app:DimensionalReduction}, the contribution of the top-form can be dualized to a potential, resulting in the theory
\begin{equation}
    S \sim \int d^dx \sqrt{g} \left( R - \frac{1}{2}(\partial \phi)^2-V(\phi)-\frac{A^2}{2}e^{-a\phi}\right),
\end{equation}
where $A$ is some positive constant. The initial potential $V(\phi)$ is modeled by an exponential, $V =C_V e^{\delta_V \phi}$. 

\begin{figure}[!htbp]
\centering
\includegraphics[width=0.6\textwidth]{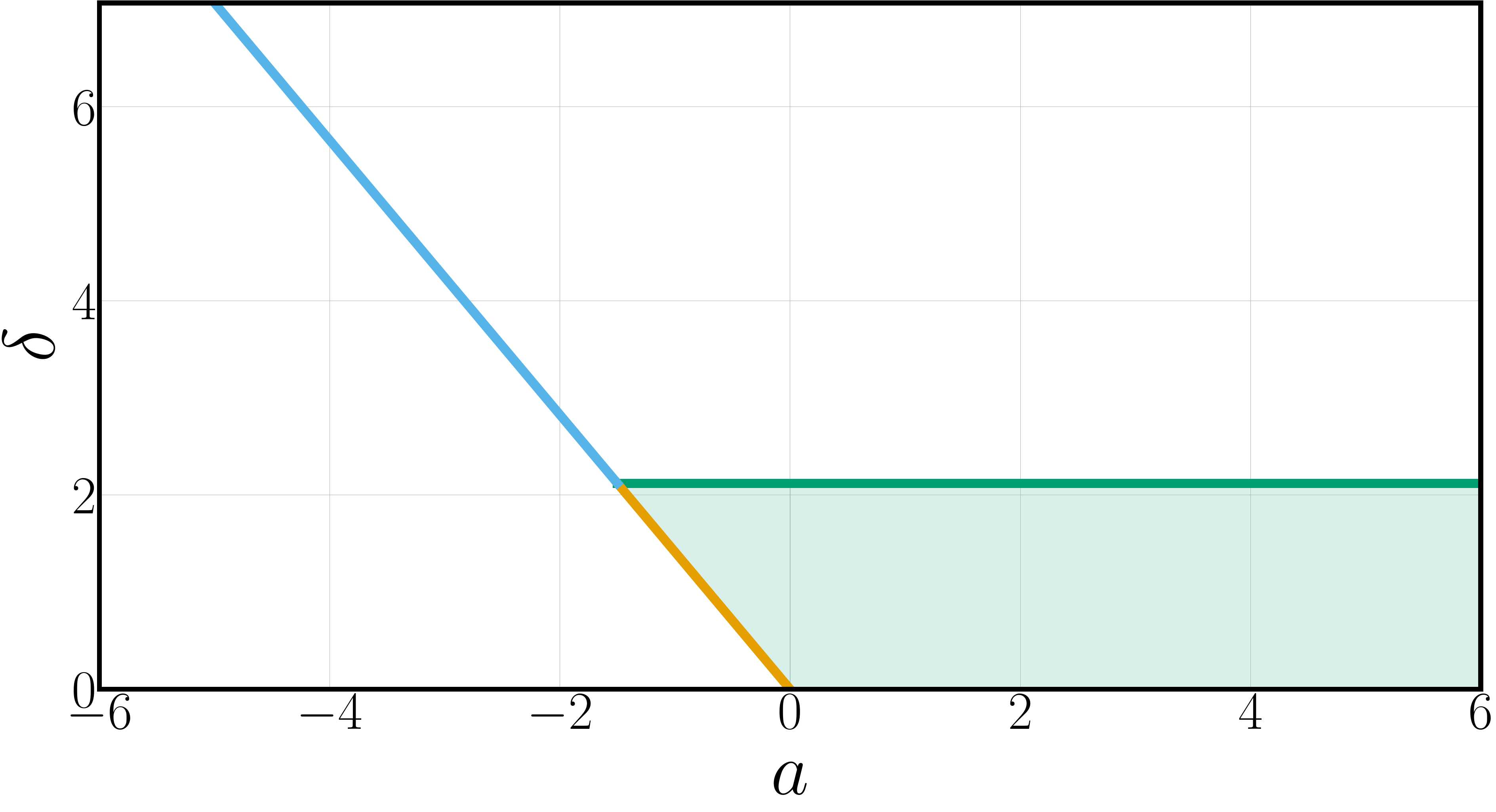}
\caption{Representation of the region compatible with $C_V<0$ in the $(a, \delta)$ plane adding a top form and assuming $d=10$. The blue line marks $\mathcal{R}^{10,1}$, the orange line $\mathcal{R}^{10,2}$, and the green region $\mathcal{R}^{10,3}$. \label{fig:DeltaAndExponentTopFormConstraints}}
\end{figure}

To find dynamical cobordism solutions we distinguish between three cases based on the relative magnitude 
of $V$ and $\frac{A^2}{2}e^{-a\phi}$:

\begin{itemize}
    \item 

Whenever the term $\frac{A^2}{2}e^{-a\phi}$ is subdominant compared to $V(\phi)$, the solution is analogous to \cite{Angius:2022aeq}. Thus, $C_V \le0$ implies
\begin{equation}
    |\delta_V| \le \sqrt{2} \sqrt{\frac{d-1}{d-2}}.
\end{equation}
The subdominance of $\frac{A^2}{2}e^{-a\phi}$ must be imposed as a consistency condition, which implies 
\begin{equation}
    \begin{aligned}
        \text{If } \delta_V>0, \quad \delta_V >-a, \\
        \text{If } \delta_V<0, \quad \delta_V <-a.
    \end{aligned}
\end{equation}
If additionally $V(\phi)$ is subdominant compared to the kinetic term, $\delta_V$ is saturated, i.e. $\delta_V=\sqrt{2}\sqrt{\frac{d-1}{d-2}}$.

\item Whenever $V(\phi)$ is subdominant compared to the top-form contribution to the potential, the solution is again analogous to \cite{Angius:2022aeq}, but now $\delta_V$ is defined by the critical exponent $\delta$ and is unrelated to that of $V(\phi)$. Then, there are two solutions
\begin{equation}
    \delta_V = -a, \qquad \delta_V = \pm \sqrt{2} \sqrt{\frac{d-1}{d-2}}. 
\end{equation}
Note that, as $A^2\ge0$, this implies that $|\delta_V| \ge \sqrt{2} \sqrt{{(d-1
)}/{(d-2)}}$, with the saturation happening at $A=0$. 

\item When $V(\phi)$ and $\frac{A^2}{2}e^{-a\phi}$ both contribute to the equations of motion, this implies $\delta_V =-a$, but now, $C_V \le0$ includes the case  $\delta_V < \sqrt{2} \sqrt{{(d-1
)}/{(d-2)}}$, which was previously not in $\mathcal{R}$. 

\end{itemize}
All in all, the region $\mathcal{R}^d$ assuming $\delta_V >0$ is given by 
\begin{equation}
    \mathcal{R}^{d}= \mathcal{R}^{d,1}\cup \mathcal{R}^{d,2} \cup \mathcal{R}^{d,3} \, 
\end{equation}
where we have defined:

\begin{subequations}
    \begin{align}
      &\mathcal{R}^{d,1}:  \delta = -\sqrt{2}a, & \text{if }& a<-\sqrt{2} \sqrt{ \frac{d-1}{d-2}}, \\
      & \mathcal{R}^{d,2}:  \delta \in \left[-a,\;2 \sqrt{ \frac{d-1}{d-2}} \right],  &\text{if }& -\sqrt{2} \sqrt{ \frac{d-1}{d-2}}<a<0, \\
      &\mathcal{R}^{d,3}: \delta \in \left[0,\;2 \sqrt{ \frac{d-1}{d-2}} \right],  &\text{if }&  0 < a.
    \end{align}
\end{subequations}
In figure \ref{fig:DeltaAndExponentTopFormConstraints} we show the allowed regions in the $(a,\delta)$-plane for $d=10$.

\section{Finite action and \texorpdfstring{$\mathcal{R}$}{R}}
\label{app:finite_action}

In this appendix we want to discuss an alternative way of establishing the allowed region, which pertains to demanding that the solution corresponds to a finite action. This is not an unreasonable demand, and has been used in the past as a guideline for the construction of solutions, see e.g., \cite{Khuri:1992ww}. We observe that in the case of the theory with only this scalar field this imposes the exact same bound as our ''Gubser-inspired'' method, while in the case of the higher form it is slightly stricter.

\subsection{Finiteness of the action in an ED theory}

Substituting the dynamical cobordism solution
\eqref{eq:DC_Solution_Scalar} in the action \eqref{eq:ActionScalarField} at leading order we get
\begin{equation}\label{eq:1USE_SubstitutedScalarAction}
    S \sim \int d^dx\; z^{\frac{2(d-1)}{(d-2)\delta_V^2}-2} \left( \frac{2d}{(d-2)\delta_V^4}\left( \delta_V^2-\frac{2(d-1)}{d-2}\right) -  a\right).
\end{equation}

We integrate between a cut-off $\Lambda$, such that $\Lambda \rightarrow 0$ corresponds to integrating the singularity, and another unspecified limit. There are three cases depending on the exponent of $z$, whose dependence in $\Lambda$ are
\begin{subequations}\label{eq:1USE_Dependences}
    \begin{align}
       \delta_V^2 &< 2\frac{d-1}{d-2}, \qquad S \sim \Lambda^p \text{ with } p>0, \label{eq:1USE_FiniteCase}\\
       \delta_V^2 &= 2\frac{d-1}{d-2}, \qquad S \sim \ln(\Lambda), \\
       \delta_V^2 &> 2\frac{d-1}{d-2}, \qquad S \sim \Lambda^p \text{ with } p<0.
    \end{align}
\end{subequations}
The only finite case in the $\Lambda \rightarrow0$ limit is \eqref{eq:1USE_FiniteCase}. We observe that the value
\begin{equation}
    \delta_{V0} = \sqrt{2\frac{d-1}{d-2}},
\end{equation}
is responsible for the behavior changes. Note that, if we impose $\delta_V=\delta_{V0}$ we have that the first term in \eqref{eq:1USE_SubstitutedScalarAction} cancels. Thus, the logarithmic dependence of \eqref{eq:1USE_Dependences} vanishes and the action is finite in this case too. Therefore, we observe that the action is finite whenever the bound \eqref{eq:DynamicalCobordismBound} is respected, coinciding with section \ref{sec:Sharpening}.

\subsection{Finiteness of the action for a higher-form field}
\label{sec:FiniteAction_HigherForm}
Now, we add a higher-form field. The dynamical cobordism solution is substituted in the action \eqref{eq:Actino_ScalarQForm} and  the necessary constraints for its finiteness are computed. We discuss explicitly the solution where both the potential and the higher form contribute to the EOMs, with the other cases constituting a simple extension.
\begin{equation}\label{eq:1USE_SubstitutedQform}
    S \sim \int d^d x \,z^{\alpha} \left( [z^2R] - \frac{2}{\delta_V^2} -C_V/c+ c^{a/\delta_V} Q^2\right) = -\int d^d x \,  \frac{2z^{\alpha}(a+\delta_V)\Xi_2}{(d-q)^2 \delta _V^2 \left(a+(q-1) \delta _V\right)},
\end{equation}
where 
\begin{equation}
    \alpha = \frac{a^2 (d-1)-a (d-3 q+2) \delta _V+(q-1) (-2 d+2 q+1) \delta _V^2+2 (q-1) (d-q)}{(d-q) \delta _V \left(a+(q-1) \delta _V\right)}.
\end{equation}
For the action to not diverge we need to impose $\zeta>-1$. However, similarly to the scalar, the case $\zeta=-1$ also results in a finite action because the term next to $1/z$ in \eqref{eq:1USE_SubstitutedQform} cancels. Then, the condition $\zeta \ge -1$ implies
\begin{equation}\label{eq:FiniteActionCondition}
    \frac{\Xi_2}{(d-q) \delta _V \left(a+(q-1) \delta _V\right)} \ge 0.
\end{equation}
In addition, we also need to impose $Q^2 \sim\Xi_0 \Xi_2 \ge0$ as a consistency condition. Assuming $\delta_V>0$ for simplicity, taking into account the subleading cases as well we find that the action is finite whenever

\begin{equation}
\label{eq:DeltaBoundsHigherForm_FiniteAction}
S <\infty:
\begin{cases} 
    \dfrac{-a\sqrt{2}}{q-1} \le \delta \le \delta_{02} & \qquad \text{if } a < -a_c \\[1.5mm]
    \dfrac{-a\sqrt{2}}{q-1} \le \delta \le \delta_0  & \qquad\text{if } -a_c \le a < 0 \\[2mm]
    \delta \le \delta_0                        & \qquad\text{if } a \ge 0 \\
    \delta = \delta_{02}                       & \qquad \forall a
\end{cases}
\end{equation}

\begin{figure} [!htbp]
\label{comparison}
\begin{subfigure}{0.49\textwidth}
\centering
\includegraphics[width=\textwidth]{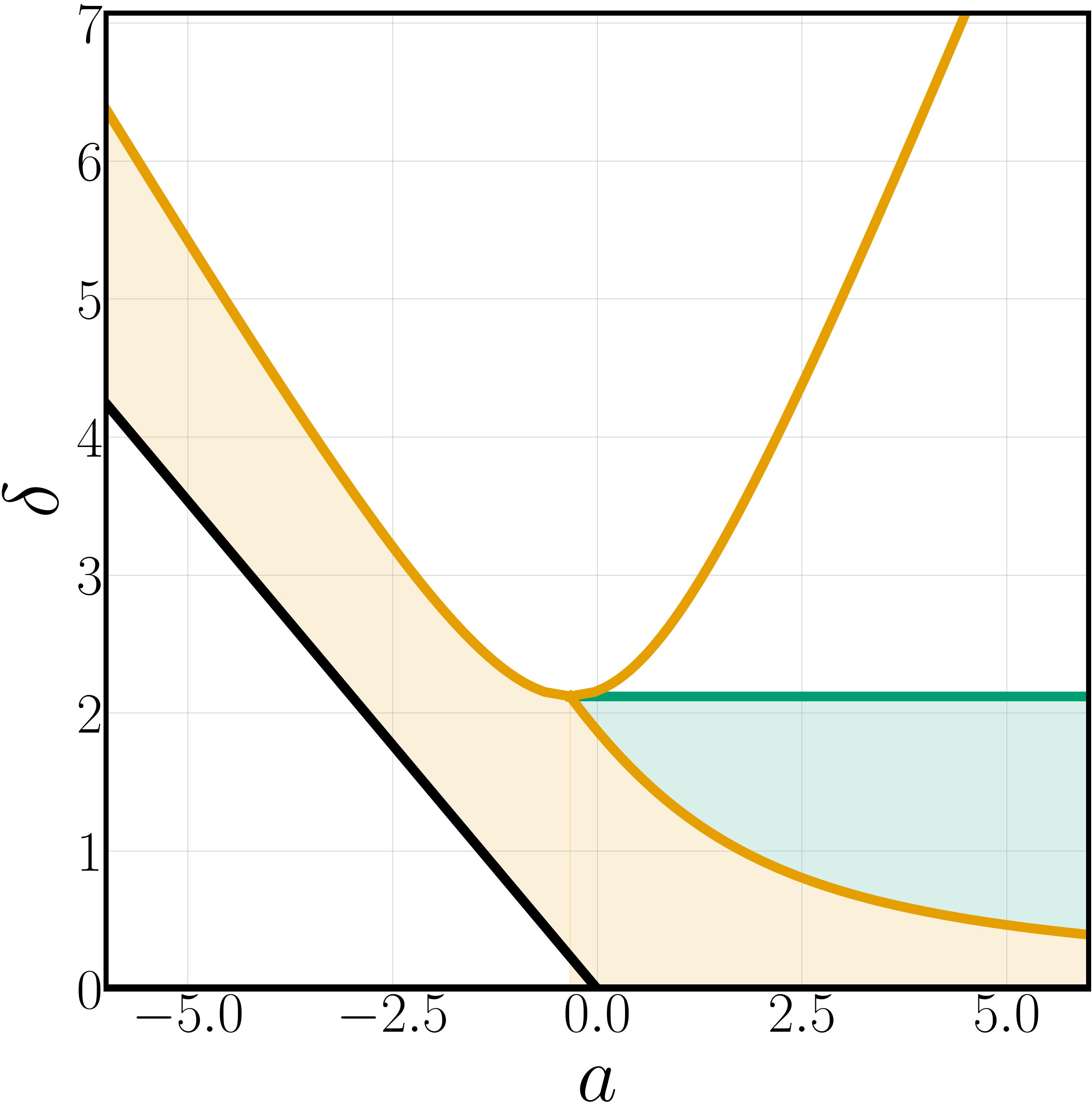}
\caption{}
\label{fig:DeltaConstraints_Action}    
\end{subfigure}
\begin{subfigure}{0.49\textwidth}
\centering
\includegraphics[width=\textwidth]{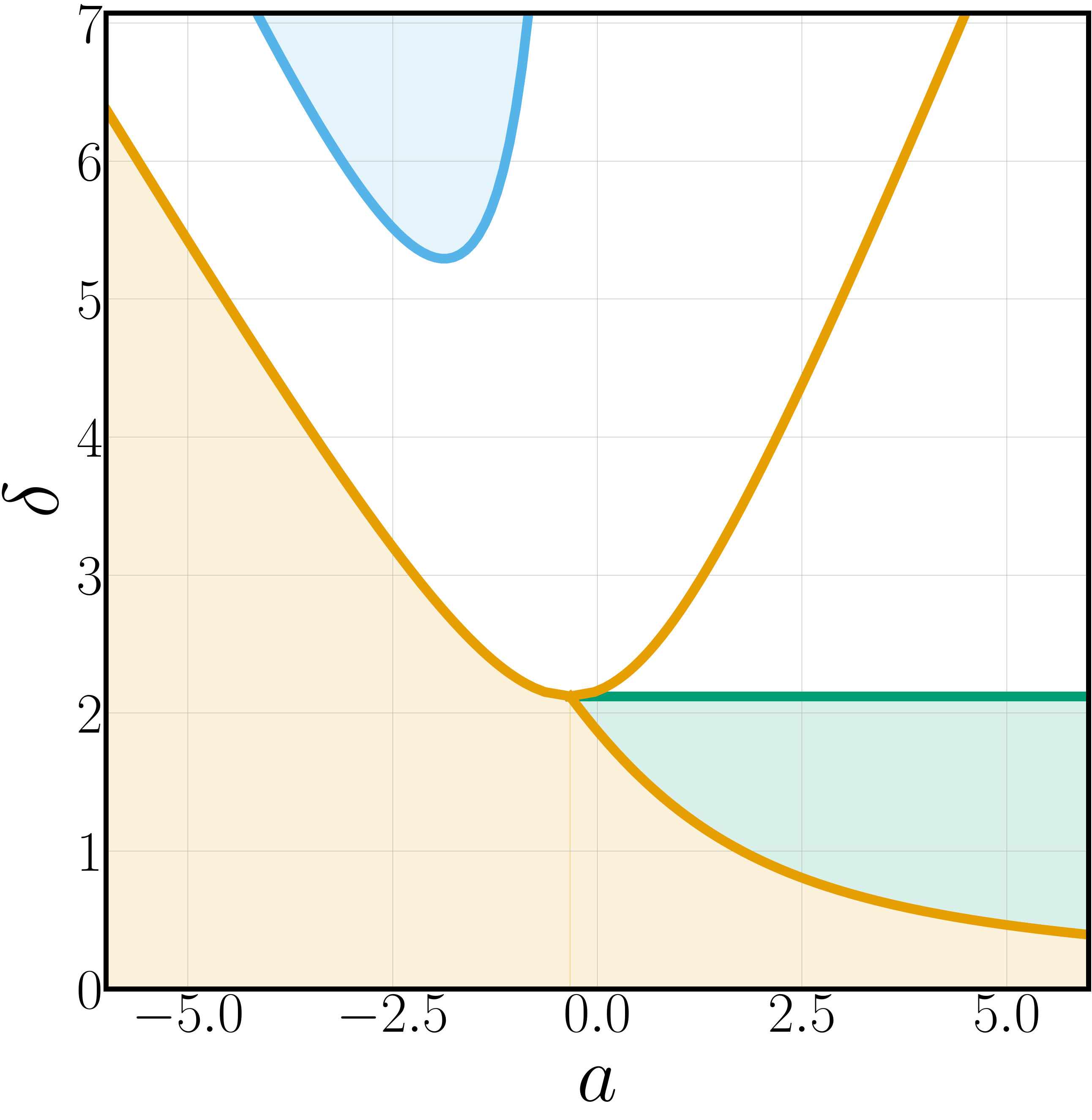}
\caption{}
\label{fig:DeltaConstraints_Potential}    
\end{subfigure}
\caption{Allowed values of $\delta$ in the $(a,\delta)$ plane for \ref{fig:DeltaConstraints_Action} requiring finiteness of the action, and \ref{fig:DeltaConstraints_Potential} $C_V\le0$ for $d=10$, $q=3$. The colors blue, orange and green are the same as in figure \ref{fig:DeltaAndExponentConstraints}, while the black line marks a new constraint given by $\delta=-a\sqrt{2}/(q-1)$.}
\end{figure}

The definition of $\delta_{02}$ and $a_c$ can be found at \eqref{eq:LimitsHigherForm}. This new set of allowed values in the parameter space for $d=10$ and $q=3$ can be observed as colored regions in figure \ref{fig:DeltaConstraints_Action} , which can be directly compared with the Gubser-inspired allowed regions in figure \ref{fig:DeltaConstraints_Potential} for the same values of $d$ and $q$.  We observe that \eqref{eq:FiniteActionCondition} does not contain the blue region $\Xi_1 \le 0$. Indeed, if we substitute the corresponding values of $\delta_V$ given by $\Xi_1 \le0$ we obtain that the action diverges. Secondly, we observe how the region $\Xi_0 \ge 0$, $\Xi_2 \ge0$ now has a new boundary, the black contour in figure \ref{fig:DeltaConstraints_Action}, given by
\begin{equation}
    \delta_V=\frac{-a}{q-1}, \qquad \delta_V=0.
\end{equation}

\clearpage
\bibliographystyle{utphys}

\bibliography{references}

@article{Breitenlohner:1982jf,
    author = "Breitenlohner, Peter and Freedman, Daniel Z.",
    title = "{Stability in Gauged Extended Supergravity}",
    reportNumber = "Print-82-0500 (MIT)",
    doi = "10.1016/0003-4916(82)90116-6",
    journal = "Annals Phys.",
    volume = "144",
    pages = "249",
    year = "1982"
}

@article{Sugimoto:1999tx,
    author = "Sugimoto, Shigeki",
    title = "{Anomaly cancellations in type I D-9 - anti-D-9 system and the USp(32) string theory}",
    eprint = "hep-th/9905159",
    archivePrefix = "arXiv",
    reportNumber = "YITP-99-25",
    doi = "10.1143/PTP.102.685",
    journal = "Prog. Theor. Phys.",
    volume = "102",
    pages = "685--699",
    year = "1999"
}

@article{Alvarez-Gaume:1986ghj,
    author = "Alvarez-Gaume, Luis and Ginsparg, Paul H. and Moore, Gregory W. and Vafa, C.",
    title = "{An O(16) x O(16) Heterotic String}",
    reportNumber = "HUTP-86/A013",
    doi = "10.1016/0370-2693(86)91524-8",
    journal = "Phys. Lett. B",
    volume = "171",
    pages = "155--162",
    year = "1986"
}

@article{Dixon:1986iz,
    author = "Dixon, Lance J. and Harvey, Jeffrey A.",
    editor = "Schellekens, B.",
    title = "{String Theories in Ten-Dimensions Without Space-Time Supersymmetry}",
    reportNumber = "PRINT-86-0244 (PRINCETON)",
    doi = "10.1016/0550-3213(86)90619-X",
    journal = "Nucl. Phys. B",
    volume = "274",
    pages = "93--105",
    year = "1986"
}

@article{Bergshoeff:1996ui,
    author = "Bergshoeff, E. and de Roo, M. and Green, Michael B. and Papadopoulos, G. and Townsend, P. K.",
    title = "{Duality of type II 7 branes and 8 branes}",
    eprint = "hep-th/9601150",
    archivePrefix = "arXiv",
    reportNumber = "DAMTP-R-95-55-REV, UG-15-95",
    doi = "10.1016/0550-3213(96)00171-X",
    journal = "Nucl. Phys. B",
    volume = "470",
    pages = "113--135",
    year = "1996"
}

@article{Romans:1985tz,
    author = "Romans, L. J.",
    editor = "Salam, A. and Sezgin, E.",
    title = "{Massive N=2a Supergravity in Ten-Dimensions}",
    reportNumber = "NSF-ITP-85-148",
    doi = "10.1016/0370-2693(86)90375-8",
    journal = "Phys. Lett. B",
    volume = "169",
    pages = "374",
    year = "1986"
}

@article{Apruzzi:2013yva,
    author = "Apruzzi, Fabio and Fazzi, Marco and Rosa, Dario and Tomasiello, Alessandro",
    title = "{All AdS$_7$ solutions of type II supergravity}",
    eprint = "1309.2949",
    archivePrefix = "arXiv",
    primaryClass = "hep-th",
    doi = "10.1007/JHEP04(2014)064",
    journal = "JHEP",
    volume = "04",
    pages = "064",
    year = "2014"
}

@article{Gubser:2000nd,
    author = "Gubser, Steven S.",
    title = "{Curvature singularities: The Good, the bad, and the naked}",
    eprint = "hep-th/0002160",
    archivePrefix = "arXiv",
    reportNumber = "PUPT-1916",
    doi = "10.4310/ATMP.2000.v4.n3.a6",
    journal = "Adv. Theor. Math. Phys.",
    volume = "4",
    pages = "679--745",
    year = "2000"
}

@article{Witten:1981gj,
    author = "Witten, Edward",
    title = "{Instability of the Kaluza-Klein Vacuum}",
    reportNumber = "PRINT-81-0441 (PRINCETON)",
    doi = "10.1016/0550-3213(82)90007-4",
    journal = "Nucl. Phys. B",
    volume = "195",
    pages = "481--492",
    year = "1982"
}

@article{Angius:2022aeq,
    author = "Angius, Roberta and Calder{\'o}n-Infante, Jos{\'e} and Delgado, Matilda and Huertas, Jes{\'u}s and Uranga, Angel M.",
    title = "{At the end of the world: Local Dynamical Cobordism}",
    eprint = "2203.11240",
    archivePrefix = "arXiv",
    primaryClass = "hep-th",
    reportNumber = "IFT-UAM/CSIC-22-31",
    doi = "10.1007/JHEP06(2022)142",
    journal = "JHEP",
    volume = "06",
    pages = "142",
    year = "2022"
}

@article{Gouteraux:2011ce,
    author = "Gouteraux, B. and Kiritsis, E.",
    title = "{Generalized Holographic Quantum Criticality at Finite Density}",
    eprint = "1107.2116",
    archivePrefix = "arXiv",
    primaryClass = "hep-th",
    doi = "10.1007/JHEP12(2011)036",
    journal = "JHEP",
    volume = "12",
    pages = "036",
    year = "2011"
}

@article{PhysRevLett20878,
  title = {Reality of the Schwarzschild Singularity},
  author = {Janis, Allen I. and Newman, Ezra T. and Winicour, Jeffrey},
  journal = {Phys. Rev. Lett.},
  volume = {20},
  issue = {16},
  pages = {878--880},
  numpages = {0},
  year = {1968},
  month = {04},
  publisher = {American Physical Society},
  doi = {10.1103/PhysRevLett.20.878},
  url = {https://link.aps.org/doi/10.1103/PhysRevLett.20.878}
}

@article{Wyman:1981bd,
    author = "Wyman, M.",
    title = "{Static Spherically Symmetric Scalar Fields in General Relativity}",
    doi = "10.1103/PhysRevD.24.839",
    journal = "Phys. Rev. D",
    volume = "24",
    pages = "839--841",
    year = "1981"
}

@article{wald1997,
    author = "Wald, Robert M.",
    title = "{Gravitational collapse and cosmic censorship}",
    eprint = "gr-qc/9710068",
    archivePrefix = "arXiv",
    reportNumber = "EFI-97-43",
    doi = "10.1007/978-94-017-0934-7_5",
    pages = "69--85",
    month = "10",
    year = "1997"
}

@article{Garfinkle:1990qj,
    author = "Garfinkle, David and Horowitz, Gary T. and Strominger, Andrew",
    title = "{Charged black holes in string theory}",
    reportNumber = "UCSB-TH-90-66",
    doi = "10.1103/PhysRevD.43.3140",
    journal = "Phys. Rev. D",
    volume = "43",
    pages = "3140",
    year = "1991",
    note = "[Erratum: Phys.Rev.D 45, 3888 (1992)]"
}

@article{Dudas:2025ubq,
    author = "Dudas, E. and Mourad, J. and Sagnotti, A.",
    title = "{Supersymmetry breaking with fields, strings and branes}",
    eprint = "2511.04367",
    archivePrefix = "arXiv",
    primaryClass = "hep-th",
    doi = "10.1016/j.physrep.2026.02.005",
    journal = "Phys. Rept.",
    volume = "1175",
    pages = "1--256",
    year = "2026"
}

@article{GIBBONS1988741,
title = {Black holes and membranes in higher-dimensional theories with dilaton fields},
journal = {Nuclear Physics B},
volume = {298},
number = {4},
pages = {741-775},
year = {1988},
issn = {0550-3213},
doi = {https://doi.org/10.1016/0550-3213(88)90006-5},
url = {https://www.sciencedirect.com/science/article/pii/0550321388900065},
author = {G.W. Gibbons and Kei-ichi Maeda},
}

@article{Freedman:1999gk,
    author = "Freedman, D. Z. and Gubser, S. S. and Pilch, K. and Warner, N. P.",
    title = "{Continuous distributions of D3-branes and gauged supergravity}",
    eprint = "hep-th/9906194",
    archivePrefix = "arXiv",
    reportNumber = "CERN-TH-99-189, HUTP-99-A029, MIT-CTP-2877, USC-99-03",
    doi = "10.1088/1126-6708/2000/07/038",
    journal = "JHEP",
    volume = "07",
    pages = "038",
    year = "2000"
}

@article{Johnson:1999qt,
    author = "Johnson, Clifford V. and Peet, Amanda W. and Polchinski, Joseph",
    title = "{Gauge theory and the excision of repulson singularities}",
    eprint = "hep-th/9911161",
    archivePrefix = "arXiv",
    reportNumber = "NSF-ITP-99-122, IASSNS-HEP-99-107, DTP-99-81",
    doi = "10.1103/PhysRevD.61.086001",
    journal = "Phys. Rev. D",
    volume = "61",
    pages = "086001",
    year = "2000"
}

@article{Kallosh:1995yz,
    author = "Kallosh, Renata and Linde, Andrei D.",
    title = "{Exact supersymmetric massive and massless white holes}",
    eprint = "hep-th/9507022",
    archivePrefix = "arXiv",
    reportNumber = "SU-ITP-95-14",
    doi = "10.1103/PhysRevD.52.7137",
    journal = "Phys. Rev. D",
    volume = "52",
    pages = "7137--7145",
    year = "1995"
}

@article{Cvetic:1995mx,
    author = "Cvetic, Mirjam and Youm, Donam",
    title = "{Singular BPS saturated states and enhanced symmetries of four-dimensional N=4 supersymmetric string vacua}",
    eprint = "hep-th/9507160",
    archivePrefix = "arXiv",
    reportNumber = "UPR-674-T, NSF-ITP-95-64",
    doi = "10.1016/0370-2693(95)01092-5",
    journal = "Phys. Lett. B",
    volume = "359",
    pages = "87--92",
    year = "1995"
}

@article{Johnson:1995bf,
    author = "Johnson, Clifford V. and Kaloper, Nemanja and Khuri, Ramzi R. and Myers, Robert C.",
    title = "{Is string theory a theory of strings?}",
    eprint = "hep-th/9509070",
    archivePrefix = "arXiv",
    reportNumber = "NSF-ITP-95-108, MCGILL-95-45, CERN-TH-95-236, PUPT-1562A, PUPT-1561",
    doi = "10.1016/0370-2693(95)01493-4",
    journal = "Phys. Lett. B",
    volume = "368",
    pages = "71--77",
    year = "1996"
}

@article{Khuri:1992ww,
    author = "Khuri, Ramzi R.",
    title = "{A Multimonopole solution in string theory}",
    eprint = "hep-th/9205051",
    archivePrefix = "arXiv",
    reportNumber = "CTP-TAMU-33-92",
    doi = "10.1016/0370-2693(92)91528-H",
    journal = "Phys. Lett. B",
    volume = "294",
    pages = "325--330",
    year = "1992"
}

@ARTICLE{Penrose:1969,
       author = {{Penrose}, Roger},
        title = "{Gravitational Collapse: the Role of General Relativity}",
      journal = {Nuovo Cimento Rivista Serie},
         year = 1969,
        month = jan,
       volume = {1},
        pages = {252},
       adsurl = {https://ui.adsabs.harvard.edu/abs/1969NCimR...1..252P}
}

@article{Maldacena:2000mw,
    author = "Maldacena, Juan Martin and Nunez, Carlos",
    editor = "Duff, Michael J. and Liu, J. T. and Lu, J.",
    title = "{Supergravity description of field theories on curved manifolds and a no go theorem}",
    eprint = "hep-th/0007018",
    archivePrefix = "arXiv",
    doi = "10.1142/S0217751X01003937",
    journal = "Int. J. Mod. Phys. A",
    volume = "16",
    pages = "822--855",
    year = "2001"
}

@article{Behrndt:1995tr,
    author = "Behrndt, Klaus",
    title = "{About a class of exact string backgrounds}",
    eprint = "hep-th/9506106",
    archivePrefix = "arXiv",
    reportNumber = "HUB-EP-95-6",
    doi = "10.1016/0550-3213(95)00506-N",
    journal = "Nucl. Phys. B",
    volume = "455",
    pages = "188--210",
    year = "1995"
}

@article{Buratti:2021yia,
    author = "Buratti, Ginevra and Delgado, Matilda and Uranga, Angel M.",
    title = "{Dynamical tadpoles, stringy cobordism, and the SM from spontaneous compactification}",
    eprint = "2104.02091",
    archivePrefix = "arXiv",
    primaryClass = "hep-th",
    doi = "10.1007/JHEP06(2021)170",
    journal = "JHEP",
    volume = "06",
    pages = "170",
    year = "2021"
}

@article{Buratti:2021fiv,
    author = "Buratti, Ginevra and Calder{\'o}n-Infante, Jos{\'e} and Delgado, Matilda and Uranga, Angel M.",
    title = "{Dynamical Cobordism and Swampland Distance Conjectures}",
    eprint = "2107.09098",
    archivePrefix = "arXiv",
    primaryClass = "hep-th",
    doi = "10.1007/JHEP10(2021)037",
    journal = "JHEP",
    volume = "10",
    pages = "037",
    year = "2021"
}

@article{Blumenhagen:2022mqw,
    author = "Blumenhagen, Ralph and Cribiori, Niccol{\`o} and Kneissl, Christian and Makridou, Andriana",
    title = "{Dynamical cobordism of a domain wall and its companion defect 7-brane}",
    eprint = "2205.09782",
    archivePrefix = "arXiv",
    primaryClass = "hep-th",
    reportNumber = "MPP-2022-57",
    doi = "10.1007/JHEP08(2022)204",
    journal = "JHEP",
    volume = "08",
    pages = "204",
    year = "2022"
}

@article{Angius:2022mgh,
    author = "Angius, Roberta and Delgado, Matilda and Uranga, Angel M.",
    title = "{Dynamical Cobordism and the beginning of time: supercritical strings and tachyon condensation}",
    eprint = "2207.13108",
    archivePrefix = "arXiv",
    primaryClass = "hep-th",
    doi = "10.1007/JHEP08(2022)285",
    journal = "JHEP",
    volume = "08",
    pages = "285",
    year = "2022"
}

@article{Blumenhagen:2023abk,
    author = "Blumenhagen, Ralph and Kneissl, Christian and Wang, Chuying",
    title = "{Dynamical Cobordism Conjecture: solutions for end-of-the-world branes}",
    eprint = "2303.03423",
    archivePrefix = "arXiv",
    primaryClass = "hep-th",
    reportNumber = "MPP-2023-33",
    doi = "10.1007/JHEP05(2023)123",
    journal = "JHEP",
    volume = "05",
    pages = "123",
    year = "2023"
}

@article{Angius:2023xtu,
    author = "Angius, Roberta and Huertas, Jesus and Uranga, Angel M.",
    title = "{Small black hole explosions}",
    eprint = "2303.15903",
    archivePrefix = "arXiv",
    primaryClass = "hep-th",
    reportNumber = "IFT-UAM/CSIC-23-31",
    doi = "10.1007/JHEP06(2023)070",
    journal = "JHEP",
    volume = "06",
    pages = "070",
    year = "2023"
}

@article{Huertas:2023syg,
    author = "Huertas, Jes{\'u}s and Uranga, Angel M.",
    title = "{Aspects of dynamical cobordism in AdS/CFT}",
    eprint = "2306.07335",
    archivePrefix = "arXiv",
    primaryClass = "hep-th",
    doi = "10.1007/JHEP08(2023)140",
    journal = "JHEP",
    volume = "08",
    pages = "140",
    year = "2023"
}

@article{Angius:2023uqk,
    author = "Angius, Roberta and Makridou, Andriana and Uranga, Angel M.",
    title = "{Intersecting end of the world branes}",
    eprint = "2312.16286",
    archivePrefix = "arXiv",
    primaryClass = "hep-th",
    doi = "10.1007/JHEP03(2024)110",
    journal = "JHEP",
    volume = "03",
    pages = "110",
    year = "2024"
}

@article{Angius:2024zjv,
    author = "Angius, Roberta",
    title = "{End of the world brane networks for infinite distance limits in CY moduli space}",
    eprint = "2404.14486",
    archivePrefix = "arXiv",
    primaryClass = "hep-th",
    reportNumber = "IFT-UAM/CSIC-24-62",
    doi = "10.1007/JHEP09(2024)178",
    journal = "JHEP",
    volume = "09",
    pages = "178",
    year = "2024"
}

@article{Huertas:2024mvy,
    author = "Huertas, Jes{\'u}s and Uranga, Angel M.",
    title = "{End of the world brane dynamics in holographic 4d $ \mathcal{N} $ = 4 SU(N) with 3d $ \mathcal{N} $ = 2 boundary conditions}",
    eprint = "2410.05368",
    archivePrefix = "arXiv",
    primaryClass = "hep-th",
    doi = "10.1007/JHEP01(2025)002",
    journal = "JHEP",
    volume = "01",
    pages = "002",
    year = "2025"
}

@article{Angius:2024pqk,
    author = "Angius, Roberta and Uranga, Angel Maria and Wang, Chuying",
    title = "{End of the world boundaries for chiral quantum gravity theories}",
    eprint = "2410.07322",
    archivePrefix = "arXiv",
    primaryClass = "hep-th",
    doi = "10.1007/JHEP03(2025)064",
    journal = "JHEP",
    volume = "03",
    pages = "064",
    year = "2025"
}

@article{Calderon-Infante:2026ymy,
    author = "Calder{\'o}n-Infante, Jos{\'e} and Cheng, Gongrui and Herr{\'a}ez, Alvaro and Van Riet, Thomas",
    title = "{End-of-the-World Singularities: The Good, the Bad, and the Heated-up}",
    eprint = "2603.18133",
    archivePrefix = "arXiv",
    primaryClass = "hep-th",
    month = "3",
    year = "2026"
}

@article{Lust:2019zwm,
    author = {L{\"u}st, Dieter and Palti, Eran and Vafa, Cumrun},
    title = "{AdS and the Swampland}",
    eprint = "1906.05225",
    archivePrefix = "arXiv",
    primaryClass = "hep-th",
    doi = "10.1016/j.physletb.2019.134867",
    journal = "Phys. Lett. B",
    volume = "797",
    pages = "134867",
    year = "2019"
}

@article{Montero:2022prj,
    author = "Montero, Miguel and Vafa, Cumrun and Valenzuela, Irene",
    title = "{The dark dimension and the Swampland}",
    eprint = "2205.12293",
    archivePrefix = "arXiv",
    primaryClass = "hep-th",
    doi = "10.1007/JHEP02(2023)022",
    journal = "JHEP",
    volume = "02",
    pages = "022",
    year = "2023"
}

@article{Kiritsis:2016kog,
    author = "Kiritsis, Elias and Nitti, Francesco and Silva Pimenta, Leandro",
    title = "{Exotic RG Flows from Holography}",
    eprint = "1611.05493",
    archivePrefix = "arXiv",
    primaryClass = "hep-th",
    doi = "10.1002/prop.201600120",
    journal = "Fortsch. Phys.",
    volume = "65",
    number = "2",
    pages = "1600120",
    year = "2017"
}

@article{Kiritsis:2025ytb,
    author = "Kiritsis, Elias and Morales-Tejera, Sergio and Rosen, Christopher",
    title = "{de Sitter versus Anti de Sitter flows and the (super)gravity landscape: Part II}",
    eprint = "2510.12373",
    archivePrefix = "arXiv",
    primaryClass = "hep-th",
    reportNumber = "CCTP-2025-10; ITCP-IPP-2020/10",
    month = "10",
    year = "2025"
}

@article{Apers:2025pon,
    author = "Apers, Fien and Montero, Miguel and Valenzuela, Irene",
    title = "{Backtracking AdS flux vacua}",
    eprint = "2506.03314",
    archivePrefix = "arXiv",
    primaryClass = "hep-th",
    reportNumber = "CERN-TH-2025-109, IFT-25-062",
    doi = "10.1007/JHEP03(2026)161",
    journal = "JHEP",
    volume = "03",
    pages = "161",
    year = "2026"
}

@article{Bedroya:2025ltj,
    author = "Bedroya, Alek and Steinhardt, Paul J.",
    title = "{Holography vs. Scale Separation}",
    eprint = "2509.25313",
    archivePrefix = "arXiv",
    primaryClass = "hep-th",
    month = "9",
    year = "2025"
}

@article{Medevielle:2023jmn,
    author = "Medevielle, Maxime and Mohaupt, Thomas",
    title = "{T-duality across non-extremal horizons}",
    eprint = "2401.00296",
    archivePrefix = "arXiv",
    primaryClass = "hep-th",
    reportNumber = "LTH 1362",
    doi = "10.1007/JHEP09(2024)116",
    journal = "JHEP",
    volume = "09",
    pages = "116",
    year = "2024"
}

@article{Dijkgraaf:2016lym,
    author = "Dijkgraaf, Robbert and Heidenreich, Ben and Jefferson, Patrick and Vafa, Cumrun",
    title = "{Negative Branes, Supergroups and the Signature of Spacetime}",
    eprint = "1603.05665",
    archivePrefix = "arXiv",
    primaryClass = "hep-th",
    doi = "10.1007/JHEP02(2018)050",
    journal = "JHEP",
    volume = "02",
    pages = "050",
    year = "2018"
}

@aunpublished{Makridou:2026ex,
    author = "Makridou, Andriana and Puga, Alejandro",
    title = "{Dynamical Cobordism and Exotic theories}",
   note = "{Manuscript in preparation}", year = "2026"
}

@article{Vafa:2005ui,
    author = "Vafa, Cumrun",
    title = "{The String landscape and the swampland}",
    eprint = "hep-th/0509212",
    archivePrefix = "arXiv",
    reportNumber = "HUTP-05-A043",
    month = "9",
    year = "2005"
}

@article{Ooguri:2006in,
    author = "Ooguri, Hirosi and Vafa, Cumrun",
    title = "{On the Geometry of the String Landscape and the Swampland}",
    eprint = "hep-th/0605264",
    archivePrefix = "arXiv",
    reportNumber = "CALT-68-2600, HUTP-06-A017",
    doi = "10.1016/j.nuclphysb.2006.10.033",
    journal = "Nucl. Phys. B",
    volume = "766",
    pages = "21--33",
    year = "2007"
}

@article{McNamara:2019rup,
    author = "McNamara, Jacob and Vafa, Cumrun",
    title = "{Cobordism Classes and the Swampland}",
    eprint = "1909.10355",
    archivePrefix = "arXiv",
    primaryClass = "hep-th",
    month = "9",
    year = "2019"
}

@article{Mohseni:2024njl,
    author = "Mohseni, Amineh and Montero, Miguel and Vafa, Cumrun and Valenzuela, Irene",
    title = "{On measuring distances in the quantum gravity landscape}",
    eprint = "2407.02705",
    archivePrefix = "arXiv",
    primaryClass = "hep-th",
    reportNumber = "IFT-24-097, CERN-TH-2024-101",
    doi = "10.1007/JHEP12(2024)168",
    journal = "JHEP",
    volume = "12",
    pages = "168",
    year = "2024"
}

@article{Debusschere:2024rmi,
    author = "Debusschere, C{\'e}dric and Tonioni, Flavio and Van Riet, Thomas",
    title = "{A distance conjecture beyond moduli?}",
    eprint = "2407.03715",
    archivePrefix = "arXiv",
    primaryClass = "hep-th",
    doi = "10.1007/JHEP03(2025)140",
    journal = "JHEP",
    volume = "03",
    pages = "140",
    year = "2025"
}

@article{Demulder:2024glx,
    author = "Demulder, Saskia and Lust, Dieter and Raml, Thomas",
    title = "{Navigating string theory field space with geometric flows}",
    eprint = "2412.10364",
    archivePrefix = "arXiv",
    primaryClass = "hep-th",
    doi = "10.1007/JHEP05(2025)030",
    journal = "JHEP",
    volume = "05",
    pages = "030",
    year = "2025"
}

@article{Basile:2023rvm,
    author = "Basile, Ivano and Montella, Carmine",
    title = "{Domain walls and distances in discrete landscapes}",
    eprint = "2309.04519",
    archivePrefix = "arXiv",
    primaryClass = "hep-th",
    doi = "10.1007/JHEP02(2024)227",
    journal = "JHEP",
    volume = "02",
    pages = "227",
    year = "2024"
}

@article{Raucci:2026fzp,
    author = "Raucci, Salvatore and Ruiz, Ignacio and Valenzuela, Irene",
    title = "{Alice in Warpland: KK modes, Warped Compactifications and the Swampland}",
    eprint = "2603.11163",
    archivePrefix = "arXiv",
    primaryClass = "hep-th",
    reportNumber = "IFT-UAM/CSIC-26-22, CERN-TH-2026-040",
    month = "3",
    year = "2026"
}

@article{Etheredge:2024tok,
    author = "Etheredge, Muldrow and Heidenreich, Ben and Rudelius, Tom and Ruiz, Ignacio and Valenzuela, Irene",
    title = "{Taxonomy of infinite distance limits}",
    eprint = "2405.20332",
    archivePrefix = "arXiv",
    primaryClass = "hep-th",
    reportNumber = "ACFI-T24-04, CERN-TH-2024-067, IFT-UAM/CSIC-23-64",
    doi = "10.1007/JHEP03(2025)213",
    journal = "JHEP",
    volume = "03",
    pages = "213",
    year = "2025"
}

@article{Etheredge:2023odp,
    author = "Etheredge, Muldrow and Heidenreich, Ben and McNamara, Jacob and Rudelius, Tom and Ruiz, Ignacio and Valenzuela, Irene",
    title = "{Running decompactification, sliding towers, and the distance conjecture}",
    eprint = "2306.16440",
    archivePrefix = "arXiv",
    primaryClass = "hep-th",
    reportNumber = "ACFI-T23-02, CERN-TH-2023-121, IFT-UAM/CSIC-23-78",
    doi = "10.1007/JHEP12(2023)182",
    journal = "JHEP",
    volume = "12",
    pages = "182",
    year = "2023"
}

@article{Etheredge:2022opl,
    author = "Etheredge, Muldrow and Heidenreich, Ben and Kaya, Sami and Qiu, Yue and Rudelius, Tom",
    title = "{Sharpening the Distance Conjecture in diverse dimensions}",
    eprint = "2206.04063",
    archivePrefix = "arXiv",
    primaryClass = "hep-th",
    reportNumber = "ACFI-T22-07",
    doi = "10.1007/JHEP12(2022)114",
    journal = "JHEP",
    volume = "12",
    pages = "114",
    year = "2022"
}

@article{Lee:2019wij,
    author = "Lee, Seung-Joo and Lerche, Wolfgang and Weigand, Timo",
    title = "{Emergent strings from infinite distance limits}",
    eprint = "1910.01135",
    archivePrefix = "arXiv",
    primaryClass = "hep-th",
    reportNumber = "CERN-TH-2019-159",
    doi = "10.1007/JHEP02(2022)190",
    journal = "JHEP",
    volume = "02",
    pages = "190",
    year = "2022"
}

@article{Blumenhagen:2000dc,
    author = "Blumenhagen, Ralph and Font, Anamaria",
    title = "{Dilaton tadpoles, warped geometries and large extra dimensions for nonsupersymmetric strings}",
    eprint = "hep-th/0011269",
    archivePrefix = "arXiv",
    reportNumber = "HUB-EP-00-55",
    doi = "10.1016/S0550-3213(01)00028-1",
    journal = "Nucl. Phys. B",
    volume = "599",
    pages = "241--254",
    year = "2001"
}

@article{Leone:2025mwo,
    author = "Leone, Giorgio and Raucci, Salvatore",
    title = "{Aspects of strings without spacetime supersymmetry}",
    eprint = "2509.24703",
    archivePrefix = "arXiv",
    primaryClass = "hep-th",
    reportNumber = "IFT-UAM/CSIC-25-100",
    doi = "10.1007/s40766-025-00078-z",
    month = "9",
    year = "2025"
}

@article{Mourad:2024dur,
    author = "Mourad, J. and Raucci, S. and Sagnotti, A.",
    title = "{Brane-like solutions and other non-supersymmetric vacua}",
    eprint = "2406.14926",
    archivePrefix = "arXiv",
    primaryClass = "hep-th",
    doi = "10.1007/JHEP10(2024)054",
    journal = "JHEP",
    volume = "10",
    pages = "054",
    year = "2024"
}

@article{Dudas:2000ff,
    author = "Dudas, E. and Mourad, J.",
    title = "{Brane solutions in strings with broken supersymmetry and dilaton tadpoles}",
    eprint = "hep-th/0004165",
    archivePrefix = "arXiv",
    reportNumber = "LPT-ORSAY-00-43, LPTM-00-25, LPT-00-43",
    doi = "10.1016/S0370-2693(00)00734-6",
    journal = "Phys. Lett. B",
    volume = "486",
    pages = "172--178",
    year = "2000"
}

@article{Raucci:2022jgw,
    author = "Raucci, Salvatore",
    title = "{On codimension-one vacua and string theory}",
    eprint = "2206.06399",
    archivePrefix = "arXiv",
    primaryClass = "hep-th",
    doi = "10.1016/j.nuclphysb.2022.116002",
    journal = "Nucl. Phys. B",
    volume = "985",
    pages = "116002",
    year = "2022"
}

@article{Schafer-Nameki:2023jdn,
    author = "Schafer-Nameki, Sakura",
    title = "{ICTP lectures on (non-)invertible generalized symmetries}",
    eprint = "2305.18296",
    archivePrefix = "arXiv",
    primaryClass = "hep-th",
    doi = "10.1016/j.physrep.2024.01.007",
    journal = "Phys. Rept.",
    volume = "1063",
    pages = "1--55",
    year = "2024"
}

@article{Gaiotto:2014kfa,
    author = "Gaiotto, Davide and Kapustin, Anton and Seiberg, Nathan and Willett, Brian",
    title = "{Generalized Global Symmetries}",
    eprint = "1412.5148",
    archivePrefix = "arXiv",
    primaryClass = "hep-th",
    doi = "10.1007/JHEP02(2015)172",
    journal = "JHEP",
    volume = "02",
    pages = "172",
    year = "2015"
}

@article{Harlow:2018tng,
    author = "Harlow, Daniel and Ooguri, Hirosi",
    title = "{Symmetries in quantum field theory and quantum gravity}",
    eprint = "1810.05338",
    archivePrefix = "arXiv",
    primaryClass = "hep-th",
    doi = "10.1007/s00220-021-04040-y",
    journal = "Commun. Math. Phys.",
    volume = "383",
    number = "3",
    pages = "1669--1804",
    year = "2021"
}

@article{Banks:2010zn,
    author = "Banks, Tom and Seiberg, Nathan",
    title = "{Symmetries and Strings in Field Theory and Gravity}",
    eprint = "1011.5120",
    archivePrefix = "arXiv",
    primaryClass = "hep-th",
    doi = "10.1103/PhysRevD.83.084019",
    journal = "Phys. Rev. D",
    volume = "83",
    pages = "084019",
    year = "2011"
}

@article{Gendler:2020dfp,
    author = "Gendler, Naomi and Valenzuela, Irene",
    title = "{Merging the weak gravity and distance conjectures using BPS extremal black holes}",
    eprint = "2004.10768",
    archivePrefix = "arXiv",
    primaryClass = "hep-th",
    doi = "10.1007/JHEP01(2021)176",
    journal = "JHEP",
    volume = "01",
    pages = "176",
    year = "2021"
}

@article{Lanza:2020qmt,
    author = "Lanza, Stefano and Marchesano, Fernando and Martucci, Luca and Valenzuela, Irene",
    title = "{Swampland Conjectures for Strings and Membranes}",
    eprint = "2006.15154",
    archivePrefix = "arXiv",
    primaryClass = "hep-th",
    doi = "10.1007/JHEP02(2021)006",
    journal = "JHEP",
    volume = "02",
    pages = "006",
    year = "2021"
}

@article{Andriot:2020lea,
    author = "Andriot, David and Cribiori, Niccol{\`o} and Erkinger, David",
    title = "{The web of swampland conjectures and the TCC bound}",
    eprint = "2004.00030",
    archivePrefix = "arXiv",
    primaryClass = "hep-th",
    doi = "10.1007/JHEP07(2020)162",
    journal = "JHEP",
    volume = "07",
    pages = "162",
    year = "2020"
}

@article{Calderon-Infante:2020dhm,
    author = "Calder{\'o}n-Infante, Jos{\'e} and Uranga, Angel M. and Valenzuela, Irene",
    title = "{The Convex Hull Swampland Distance Conjecture and Bounds on Non-geodesics}",
    eprint = "2012.00034",
    archivePrefix = "arXiv",
    primaryClass = "hep-th",
    reportNumber = "IFT-UAM/CSIC-20-169",
    doi = "10.1007/JHEP03(2021)299",
    journal = "JHEP",
    volume = "03",
    pages = "299",
    year = "2021"
}

@article{Heckman:2025wqd,
    author = "Heckman, Jonathan J. and McNamara, Jacob and Parra-Martinez, Julio and Torres, Ethan",
    title = "{GSO Defects: IIA/IIB Walls and the Surprisingly Stable $\mathrm{R}7$-Brane}",
    eprint = "2507.21210",
    archivePrefix = "arXiv",
    primaryClass = "hep-th",
    reportNumber = "CERN-TH-2025-136",
    month = "7",
    year = "2025"
}

@article{Dierigl:2023jdp,
    author = "Dierigl, Markus and Heckman, Jonathan J. and Montero, Miguel and Torres, Ethan",
    title = "{R7-branes as charge conjugation operators}",
    eprint = "2305.05689",
    archivePrefix = "arXiv",
    primaryClass = "hep-th",
    reportNumber = "LMU-ASC 16/23, IFT-UAM/CSIC-23-50",
    doi = "10.1103/PhysRevD.109.046004",
    journal = "Phys. Rev. D",
    volume = "109",
    number = "4",
    pages = "046004",
    year = "2024"
}

@article{Dierigl:2022reg,
    author = "Dierigl, Markus and Heckman, Jonathan J. and Montero, Miguel and Torres, Ethan",
    title = "{IIB string theory explored: Reflection 7-branes}",
    eprint = "2212.05077",
    archivePrefix = "arXiv",
    primaryClass = "hep-th",
    reportNumber = "LMU-ASC 31/22",
    doi = "10.1103/PhysRevD.107.086015",
    journal = "Phys. Rev. D",
    volume = "107",
    number = "8",
    pages = "086015",
    year = "2023"
}

@article{Chakrabhavi:2025bfi,
    author = "Chakrabhavi, Vivek and Debray, Arun and Dierigl, Markus and Heckman, Jonathan J.",
    title = "{Exploring Pintopia: Reflection Branes, Bordisms, and U-Dualities}",
    eprint = "2509.03573",
    archivePrefix = "arXiv",
    primaryClass = "hep-th",
    reportNumber = "CERN-TH-2025-180",
    month = "9",
    year = "2025"
}

@article{Nevoa:2025xiq,
    author = "Nevoa, Vinicius and Raman, Sanjay and Vafa, Cumrun",
    title = "{Elementary Constituents Conjecture}",
    eprint = "2511.13813",
    archivePrefix = "arXiv",
    primaryClass = "hep-th",
    month = "11",
    year = "2025"
}

@article{Cavusoglu:2026xiv,
    author = "{\c{C}}avu{\c{s}}o{\u{g}}lu, Atakan and Cveti{\v{c}}, Mirjam and Heckman, Jonathan J. and Kuntz, Jeffrey and Murdia, Chitraang",
    title = "{Gravitational Background of Alice-Vortices and R7-Branes}",
    eprint = "2602.13196",
    archivePrefix = "arXiv",
    primaryClass = "hep-th",
    month = "2",
    year = "2026"
}

@article{Hamada:2025duq,
    author = "Hamada, Yu and Hamada, Yuta and Kimura, Hayate",
    title = "{Black string in the standard model}",
    eprint = "2501.05678",
    archivePrefix = "arXiv",
    primaryClass = "hep-th",
    reportNumber = "KEK-TH-2679, DESY-24-222",
    doi = "10.1103/x3sh-45j7",
    journal = "Phys. Rev. D",
    volume = "111",
    number = "12",
    pages = "125009",
    year = "2025"
}

@article{Hassfeld:2023kpu,
    author = "Hassfeld, Bjoern and Hebecker, Arthur and Walcher, Johannes",
    title = "{Cobordism and bubbles of anything in the string landscape}",
    eprint = "2310.06021",
    archivePrefix = "arXiv",
    primaryClass = "hep-th",
    doi = "10.1007/JHEP02(2024)127",
    journal = "JHEP",
    volume = "02",
    pages = "127",
    year = "2024"
}

@article{Muntz:2025ltu,
    author = "Muntz, Benjamin and Padilla, Antonio and Saffin, Paul M.",
    title = "{Do we live on the End of the World?}",
    eprint = "2411.05912",
    archivePrefix = "arXiv",
    primaryClass = "hep-th",
    doi = "10.1007/JHEP05(2025)006",
    journal = "JHEP",
    volume = "05",
    pages = "006",
    year = "2025"
}

@article{Gao:2024lrb,
    author = "Gao, Changjun and Qiu, Jianhui",
    title = "{From the Janis{\textendash}Newman{\textendash}Winicour Naked Singularities to the Einstein{\textendash}Maxwell Phantom Wormholes}",
    eprint = "2407.14184",
    archivePrefix = "arXiv",
    primaryClass = "gr-qc",
    doi = "10.3390/universe10080328",
    journal = "Universe",
    volume = "10",
    number = "8",
    pages = "328",
    year = "2024"
}

@article{Horowitz1994ExtremalBH,
  title = {Extremal Black Holes as Exact String Solutions},
  author = {Horowitz, Gary T. and Tseytlin, A. A.},
  journal = {Phys. Rev. Lett.},
  volume = {73},
  issue = {25},
  pages = {3351--3354},
  numpages = {0},
  year = {1994},
  month = {12},
  publisher = {American Physical Society},
  doi = {10.1103/PhysRevLett.73.3351},
  url = {https://link.aps.org/doi/10.1103/PhysRevLett.73.3351}
}

@article{Horowitz:1995ta,
    author = "Horowitz, Gary T. and Myers, Robert C.",
    title = "{The value of singularities}",
    eprint = "gr-qc/9503062",
    archivePrefix = "arXiv",
    reportNumber = "UCSBTH-95-6, MCGILL-95-20",
    doi = "10.1007/BF02113073",
    journal = "Gen. Rel. Grav.",
    volume = "27",
    pages = "915--919",
    year = "1995"
}

@article{Klebanov:2000nc,
    author = "Klebanov, Igor R. and Tseytlin, Arkady A.",
    title = "{Gravity duals of supersymmetric SU(N) x SU(N+M) gauge theories}",
    eprint = "hep-th/0002159",
    archivePrefix = "arXiv",
    reportNumber = "PUPT-1919, OHSTPY-HEP-T-00-002",
    doi = "10.1016/S0550-3213(00)00206-6",
    journal = "Nucl. Phys. B",
    volume = "578",
    pages = "123--138",
    year = "2000"
}

@article{Klebanov:2000hb,
    author = "Klebanov, Igor R. and Strassler, Matthew J.",
    title = "{Supergravity and a confining gauge theory: Duality cascades and chi SB resolution of naked singularities}",
    eprint = "hep-th/0007191",
    archivePrefix = "arXiv",
    reportNumber = "IASSNS-HEP-00-56, PUPT-1944",
    doi = "10.1088/1126-6708/2000/08/052",
    journal = "JHEP",
    volume = "08",
    pages = "052",
    year = "2000"
}

@article{Altavista:2026edv,
    author = "Altavista, Chiara and Anastasi, Edoardo and Angius, Roberta and Uranga, Angel M.",
    title = "{The Art of Branching: Cobordism Junctions of 10d String Theories}",
    eprint = "2603.24667",
    archivePrefix = "arXiv",
    primaryClass = "hep-th",
    month = "3",
    year = "2026"
}

@article{Raamsdonk:2020tin,
    author = "Raamsdonk, Mark Van and Waddell, Chris",
    title = "{Holographic and localization calculations of boundary F for $ \mathcal{N} $ = 4 SUSY Yang-Mills theory}",
    eprint = "2010.14520",
    archivePrefix = "arXiv",
    primaryClass = "hep-th",
    doi = "10.1007/JHEP02(2021)222",
    journal = "JHEP",
    volume = "02",
    pages = "222",
    year = "2021"
}

@article{VanRaamsdonk:2021duo,
    author = "Van Raamsdonk, Mark and Waddell, Chris",
    title = "{Finding AdS$^{5}${\texttimes} S$^{5}$ in 2+1 dimensional SCFT physics}",
    eprint = "2109.04479",
    archivePrefix = "arXiv",
    primaryClass = "hep-th",
    doi = "10.1007/JHEP11(2021)145",
    journal = "JHEP",
    volume = "11",
    pages = "145",
    year = "2021"
}

@article{Debray:2023yrs,
    author = "Debray, Arun and Dierigl, Markus and Heckman, Jonathan J. and Montero, Miguel",
    title = "{The Chronicles of IIBordia: Dualities, Bordisms, and the Swampland}",
    eprint = "2302.00007",
    archivePrefix = "arXiv",
    primaryClass = "hep-th",
    reportNumber = "LMU-ASC 06/23, IFT-UAM/CSIC-23-7",
    month = "1",
    year = "2023"
}

@article{Basile:2023knk,
    author = "Basile, Ivano and Debray, Arun and Delgado, Matilda and Montero, Miguel",
    title = "{Global anomalies {\&} bordism of non-supersymmetric strings}",
    eprint = "2310.06895",
    archivePrefix = "arXiv",
    primaryClass = "hep-th",
    reportNumber = "IFT-23-129",
    doi = "10.1007/JHEP02(2024)092",
    journal = "JHEP",
    volume = "02",
    pages = "092",
    year = "2024"
}

@article{Braeger:2025kra,
    author = "Braeger, Noah and Debray, Arun and Dierigl, Markus and Heckman, Jonathan J. and Montero, Miguel",
    title = "{Cobordism Utopia: U-Dualities, Bordisms, and the Swampland}",
    eprint = "2505.15885",
    archivePrefix = "arXiv",
    primaryClass = "hep-th",
    reportNumber = "CERN-TH-2025-103, IFT-025-51",
    month = "5",
    year = "2025"
}

@article{Blumenhagen:2022bvh,
    author = "Blumenhagen, Ralph and Cribiori, Niccol{\`o} and Kneissl, Christian and Makridou, Andriana",
    title = "{Dimensional Reduction of Cobordism and K-theory}",
    eprint = "2208.01656",
    archivePrefix = "arXiv",
    primaryClass = "hep-th",
    reportNumber = "MPP-2022-95",
    doi = "10.1007/JHEP03(2023)181",
    journal = "JHEP",
    volume = "03",
    pages = "181",
    year = "2023"
}

@article{Kaidi:2023tqo,
    author = "Kaidi, Justin and Ohmori, Kantaro and Tachikawa, Yuji and Yonekura, Kazuya",
    title = "{Nonsupersymmetric Heterotic Branes}",
    eprint = "2303.17623",
    archivePrefix = "arXiv",
    primaryClass = "hep-th",
    doi = "10.1103/PhysRevLett.131.121601",
    journal = "Phys. Rev. Lett.",
    volume = "131",
    number = "12",
    pages = "121601",
    year = "2023"
}

@article{Kaidi:2024cbx,
    author = "Kaidi, Justin and Tachikawa, Yuji and Yonekura, Kazuya",
    title = "{On non-supersymmetric heterotic branes}",
    eprint = "2411.04344",
    archivePrefix = "arXiv",
    primaryClass = "hep-th",
    reportNumber = "TU-1248, KYUSHU-HET-294",
    doi = "10.1007/JHEP03(2025)211",
    journal = "JHEP",
    volume = "03",
    pages = "211",
    year = "2025"
}

@article{Hassfeld:2025hjx,
    author = "Hassfeld, Bjoern and Hebecker, Arthur and Schiller, Daniel",
    title = "{Localized gravity, de Sitter, and the Horizon Criterion}",
    eprint = "2505.07934",
    archivePrefix = "arXiv",
    primaryClass = "hep-th",
    doi = "10.1007/JHEP10(2025)151",
    journal = "JHEP",
    volume = "10",
    pages = "151",
    year = "2025"
}

@article{Kneissl:2024zox,
    author = "Kneissl, Christian",
    title = "{Spin cobordism and the gauge group of type I/heterotic string theory}",
    eprint = "2407.20333",
    archivePrefix = "arXiv",
    primaryClass = "hep-th",
    reportNumber = "MPP-2024-159",
    doi = "10.1007/JHEP01(2025)181",
    journal = "JHEP",
    volume = "01",
    pages = "181",
    year = "2025"
}

@article{Lanza:2021udy,
    author = "Lanza, Stefano and Marchesano, Fernando and Martucci, Luca and Valenzuela, Irene",
    title = "{The EFT stringy viewpoint on large distances}",
    eprint = "2104.05726",
    archivePrefix = "arXiv",
    primaryClass = "hep-th",
    doi = "10.1007/JHEP09(2021)197",
    journal = "JHEP",
    volume = "09",
    pages = "197",
    year = "2021"
}

@article{Coudarchet:2023mfs,
    author = "Coudarchet, Thibaut",
    title = "{Hiding the extra dimensions: A review on scale separation in string theory}",
    eprint = "2311.12105",
    archivePrefix = "arXiv",
    primaryClass = "hep-th",
    doi = "10.1016/j.physrep.2024.02.003",
    journal = "Phys. Rept.",
    volume = "1064",
    pages = "1--28",
    year = "2024"
}

@article{DeWolfe:2005uu,
    author = "DeWolfe, Oliver and Giryavets, Alexander and Kachru, Shamit and Taylor, Washington",
    title = "{Type IIA moduli stabilization}",
    eprint = "hep-th/0505160",
    archivePrefix = "arXiv",
    reportNumber = "MIT-CTP-3640, PUPT-2161, SU-ITP-05-16, SLAC-PUB-11153",
    doi = "10.1088/1126-6708/2005/07/066",
    journal = "JHEP",
    volume = "07",
    pages = "066",
    year = "2005"
}

@article{Demulder:2026cfo,
    author = "Demulder, Saskia and Lust, Dieter and Montella, Carmine and Raml, Thomas",
    title = "{Optimal paths across potentials on scalar field space}",
    eprint = "2604.24843",
    archivePrefix = "arXiv",
    primaryClass = "hep-th",
    month = "4",
    year = "2026"
}

@article{Hellerman:2010dv,
    author = "Hellerman, Simeon and Kleban, Matthew",
    title = "{Dynamical Cobordisms in General Relativity and String Theory}",
    eprint = "1009.3277",
    archivePrefix = "arXiv",
    primaryClass = "hep-th",
    reportNumber = "IPMU-10-0156",
    doi = "10.1007/JHEP02(2011)022",
    journal = "JHEP",
    volume = "02",
    pages = "022",
    year = "2011"
}

\end{document}